\documentclass[amsmath,amssymb,aps,superscriptaddress]{revtex4-2}

\usepackage{hyperref}
\usepackage{graphicx,epstopdf,pgfplots,caption,subcaption}

\usepackage{mathtools}
\mathtoolsset{showonlyrefs=false}

\renewcommand{\vec}[1]{\boldsymbol{#1}}
\newcommand{\p}[0]{\partial}

\newcommand{\bmth}[1]{\mbox{\boldmath $#1$}}
\newcommand{\grad}{\bmth{\nabla}}
\newcommand{\Ca}{\textrm{Ca}}
\newcommand{\Comp}{\mathcal{C}}

\newcommand{\res}{\mathrm{res}}
\newcommand{\atm}{\textrm{atm}}
\newcommand{\RRC}[1]{\lambda(#1)}

\begin{document}
	
\title{Compression-driven viscous fingering in a radial Hele-Shaw cell}
\author{Callum Cuttle}\thanks{These authors contributed equally to this work.}
\author{Liam C. Morrow}\thanks{These authors contributed equally to this work.}
\author{Christopher W. MacMinn}
\email{christopher.macminn@eng.ox.ac.uk}
\affiliation{Department of Engineering Science,\\
	University of Oxford, Oxford, OX1 3PJ, UK \\
}
	
\begin{abstract}
The displacement of a viscous liquid by a gas within a Hele-Shaw cell is a classical problem. The gas--liquid interface is hydrodynamically unstable, forming striking finger-like patterns that have attracted research interest for decades. Generally, both the gas and liquid phases are taken to be incompressible, with the capillary number being the key parameter that determines the severity of the instability. Here, we consider a radially outward displacement driven by the steady compression of a gas reservoir. The associated gas-injection rate is then unsteady due to the compressibility of the gas. We identify a second nondimensional parameter, the compressibility number, that plays a strong role in the development of the fingering pattern. We use an axisymmetric model to study the impact of compressibility number on the unsteady evolution of injection rate and gas pressure. We use linear stability analysis to show that increasing the compressibility number delays the onset of finger development relative to the corresponding incompressible case. Finally, we present and compare a series of experiments and fully nonlinear simulations over a broad range of capillary and compressibility numbers. These results show that increasing the compressibility number systematically decreases the severity of the fingering pattern at high capillary number. Our results provide an unprecedented comparison of experiments with simulations for viscous fingering, a comprehensive understanding of the role of compressibility in unstable gas--liquid displacement flows, and insight into a new mechanism for controlling the development of fingering patterns.
\end{abstract}
	
\date{\today}
	
\maketitle

\section{Introduction}

When a fluid is displaced from a Hele-Shaw cell or porous medium by the injection of a less viscous fluid, the fluid-fluid interface is hydrodynamically unstable and tends to deform into complex, branched structures~\citep{engelberts-3wpc-1951,Hill1952,Chuoke1959,Saffman1958,Homsy1987}. This classical viscous-fingering instability has been extensively studied as an archetype of interfacial pattern formation \citep{Paterson1981,Couder1986,Casademunt2004} and for its relevance to enhanced oil recovery \citep{engelberts-3wpc-1951,vanmeurs-transaime-1957,vanmeurs-transaime-1958,Chuoke1959,lake-prenticehall-1989}, for which fingering poses major obstacles. Modern applications include the operation of fuel cells~\citep{Lee2019,Mortazavi2020}, the remediation of groundwater contamination~\citep{Clayton1998,Hu2010}, and the subsurface sequestration of CO$_2$~\citep{cinar-spej-2009,Wang2013} or storage of hydrogen~\citep{Paterson1983}. Whether fingering is advantageous or problematic, the key concern in all applications has been understanding the mechanisms that influence the development of the fingering pattern, which is driven by viscous forces in the fluids and opposed by capillary forces at the interface. The capillary number $\Ca$ \citep{Park1985}, which measures the relative scales of these forces, is therefore the key control parameter. Recent studies have considered a variety of perturbations to classical viscous fingering, such as  imposing a time-dependent injection rate \citep{Dias2012,Morrow2019}, replacing one of the rigid plates with an elastic membrane or slab~\citep{PihlerPuzovic2012,Peng2022}, varying the gap between the plates as the gas is injected \citep{Zheng2015,Vaquero2019}, using a tapered (rather than uniform) flow cell \citep{AlHousseiny2012,Bongrand2018}, and the application of external electric fields \citep{Gao2019}. Here, we examine a simple but previously overlooked mechanism in classical gas-driven viscous fingering: the compression of the injected gas.

Gas-driven displacement of liquid is a common practical and experimental scenario that is highly susceptible to viscous fingering. In addition to generally being much less viscous than liquids, gases are also much more compressible than liquids; as a result, they will undergo some amount of spring-like volumetric compression under the typical viscous pressures of displacement flows. Such compression-driven flows can exhibit unsteady flow rates \citep{Sandnes2012, Lai2018, Cuttle2023a}, which are known to exert a fundamental influence on pattern-forming processes~\citep{Sandnes2011}. Mathematical models of gas--liquid displacement in a Hele-Shaw cell typically assume that the gas is incompressible, whereas experimentalists often take pains to avoid or ignore gas compression, such as by withdrawing the liquid at the outlet instead of injecting the gas at the inlet~\citep{Park1984b,Tabeling1986,KopfSill1987,Tabeling1987,Moore2002,Ristroph2006}, performing analysis based on the instantaneous interface velocity \citep{Maxworthy1989,Bongrand2018}, or constraining themselves to low injection rates~\citep{Martyushev2009}. Alternatively, one may inject a low-viscosity liquid rather than a gas, such as injecting water into oil \citep{Chen1987,Zhao2016}. However, the ability to entirely neglect the viscosity of the gas phase makes gas-driven displacement flows especially tractable to analytical and numerical analysis and many applications inherently involve gases. As such, compressibility is typically considered to be a necessary complication in studies of viscous fingering, but the impact of compressibility on displacement flows and viscous fingering in Hele-Shaw cells has not previously been considered in any detail.

Here, we use a combination of mathematical modelling, simulations, and experiments to elucidate the role of gas compression during gas-driven viscous fingering. We show that the unsteady flow rates observed in experiments and simulations can be rationalised by a simple axisymmetric model that couples the compression of an ideal gas to laminar viscous flow of liquid. Within the context of this simplified model, the problem is controlled by a single dimensionless compressibility number $\Comp$ that compares the rate of viscous depressurisation to the rate of compressive pressurisation. We find that $\Comp$ also plays an important role in experiments and simulations involving fingering, dictating both the volumetric growth rate and the evolution of pressure. We examine the dynamical-systems framework that underpins the basic compression-driven displacement flow in the simplified model, identifying the emergence of two distinct dynamical regimes that are analogous to those recently described for compression-driven displacement in a capillary tube \citep{Cuttle2023a}. A linear stability analysis of this axisymmetric model suggests that compressibility can significantly delay the onset of viscous fingering at high $\Ca$, via the low initial injection rates of the compression-driven flow. This prediction is confirmed in our experiments and nonlinear simulations, where we observe a systematic delay in the onset of fingering to larger radii as $\Comp$ is increased. We further show that increasing $\Comp$ is as effective in delaying onset as decreasing $\Ca$, pointing to compressibility as a powerful new control parameter for gas-driven viscous fingering. Our study also provides an unusually thorough comparison between simulations and experiments of viscous fingering. Previous studies have made quantitative comparisons between experimental results and simulations over a relatively small portion of the parameter space \cite{Li2009,Dias2012,Morrow2019,Zhao2020,Morrow2023,Oliveira2023}. The present study is the most extensive comparison of its type, featuring quantitative examination of the fingering patterns, the time-dependent injection rate, gas pressure, and the onset and development of the instability. This work therefore provides a rare opportunity to validate a numerical realisation of viscous fingering directly against a dedicated set of experiments.
	
\begin{figure}
	\centering
	\includegraphics[width=0.5\linewidth]{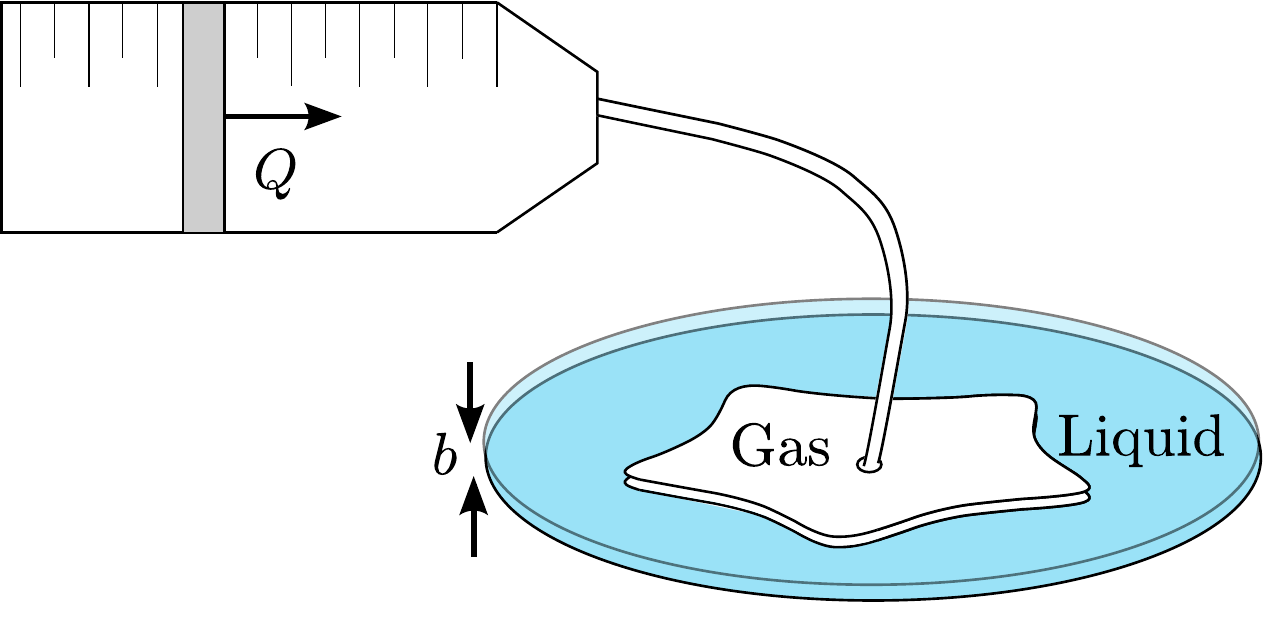}
	\caption{Schematic diagram of the fingering problem considered here. A reservoir of gas (white) is compressed at a constant volumetric rate $Q$ in order to displace liquid (blue) from a Hele-Shaw cell comprising the uniform gap $b$ between two plates.}
	\label{fig:schematic}
\end{figure}
	
\section{Mathematical Model}

\subsection{Full quasi-2D model}\label{sec:fullModel}

\begin{subequations}
	
	We consider the system shown schematically in Figure~\ref{fig:schematic}, comprising a rigid, circular Hele--Shaw cell of radius $R_c$ and gap thickness $b$, initially filled with an incompressible liquid of dynamic viscosity $\mu$. Gas is injected through a hole in the centre of the cell by compressing an attached gas reservoir of initial volume $V_{\res}(0)$ at a constant volumetric rate $Q$. We take the gas and liquid phases to be immiscible. We ignore viscous pressure gradients within the gas phase, such that the gas pressure $p_g(t)$ is spatially uniform.
	
	Within the domain $\Omega$ of the liquid phase, we consider a quasi two--dimensional (quasi-2D) model of Hele--Shaw flow that relates the gap-averaged velocity of the liquid $\vec{v}$ to its pressure $p$ via
	\begin{alignat}{3}
		\vec{v} &= -\frac{b^2}{12 \mu} \grad p, \qquad &\boldsymbol{x} &\in \mathbb{R}^2 \backslash \Omega(t). \label{eq:Darcy}
	\end{alignat}
	Further, the incompressibility of the liquid phase requires that $\grad \cdot \vec{v} = 0$ and, hence, that
	\begin{alignat}{3}
		\nabla^2 p &= 0, \qquad &\boldsymbol{x} &\in \mathbb{R}^2 \backslash \Omega(t). \label{eq:Laplace}
	\end{alignat}	
	Following \citet{Peng2015}, we impose the kinematic and dynamic conditions at the moving interface $\p \Omega(t)$:
	\begin{alignat}{3}
		v_n &= - \frac{1}{1 - f_1}\frac{b^2}{12 \mu} \grad p \cdot \vec{n},   \qquad \qquad     &\boldsymbol{x} &\in \partial \Omega(t), \label{eq:Kinematic}\\
		p &=  p_g(t)-\gamma \left( \frac{\pi}{4} \kappa + \frac{2 f_2}{b} \right) , \qquad \qquad  &\boldsymbol{x} &\in \partial \Omega(t), \label{eq:Dynamic}	
	\end{alignat}
	where $v_n$ is the normal velocity of the interface, $\vec{n}$ is the unit normal to the interface, $\gamma$ is the interfacial tension, and $\kappa$ is the signed in-plane curvature of the interface. The kinematic boundary condition Eq.~\eqref{eq:Kinematic} relates the velocity of the interface to the velocity of the liquid. The dynamic boundary condition Eq.~\eqref{eq:Dynamic} models the capillary pressure via the Young--Laplace equation. Both of these boundary conditions are influenced by the presence of thin residual films of liquid that coat the walls in the gas region~\citep{Park1984a}. Following \citet{Peng2015}, we model these films via the empirical functions
	\begin{align}
		f_1 = \frac{|v_n \eta / \gamma|^{2/3}}{0.76 + 2.16|v_n \eta / \gamma|^{2/3}}, \quad \mathrm{and} \quad f_2 = 1 + 1.59 |v_n \eta / \gamma| + \frac{|v_n \eta / \gamma|^{2/3}}{0.26 + 1.48|v_n \eta / \gamma|^{2/3} },\label{eq:films1}
	\end{align}
	which were derived by fitting to simulations of viscous fingering in a rigid Hele-Shaw channel \cite{Reinelt1985}. We assume here and going forward that the liquid is perfectly wetting to the cell walls. The total thickness of the films $f_1 b$ depends on the normal velocity of the interface at the instant of deposition, and we assume that the films are static after deposition on the timescale of the flow, such that the film thickness at each point in space is constant after the interface has passed. The function $f_2$ captures the modified dynamic boundary condition due to capillary and viscous stresses at the interface. We take the gas to be a fixed mass of ideal gas and the process to be isothermal, meaning that temperature changes due to compression and expansion equilibrate rapidly with the environment. In practice, the compression of the gas will not be perfectly isothermal due to the timescale of thermal diffusion through the glass walls of the syringe. However, our data suggests that an isothermal model is suitable for our experiments, as discussed in Appendix~\ref{app:adiabatic}. Pressure and volume are then related by Boyle's law,
	\begin{align}
		p_g(t) V_g(t) &= p_g(0) V_g(0),
	\end{align}
	where $V_g(t)=V_b(t)+V_\res(t)$ is the total volume of gas in the system, with $V_b$ the volume of gas in the cell (accounting for thin films) and $V_\res$ the volume of the gas reservoir. Due to the imposed steady compression rate $Q$, the volume of gas in the reservoir is $V_{\res}(t) = V_{\res}(0) - Qt$, such that the pressure of the gas is
	\begin{align}
		p_g(t) &= \frac{V_g(0)}{V_g(0) + V_b -V_b(0) - Qt} \left[ p_{\atm} + \gamma \left(  \frac{\pi}{4r_0} + \frac{2}{b} \right)   \right] ,\label{eq:gasPress}
	\end{align}
	where the term in square brackets is the initial gas pressure $p_g(0)$, which balances the Laplace pressure jump across the interface via Eq.~\eqref{eq:Dynamic} for $p(0)=p_\atm$. We have taken the system to be initially at rest, with a small circular bubble of gas of radius $r_0$ centred on the origin (see \S~\ref{sec:NumSche}) such that the initial in-plane curvature of the interface is $1/r_0$ and $V_b(0)=\pi r_0^2 b$. While pressure and volume must be related in terms of absolute pressure, we will find it instructive to consider gauge pressures, measured relative to atmospheric pressure. The gauge gas pressure $\Delta p_g=p_g-p_\atm$ can be written as
	\begin{align}\label{eq:gaugeGasPress}
		\Delta p_g(t) &= \frac{p_\atm V_g^{-1}(0)(Qt+V_{b0}-V_b) +  \gamma \left(  \frac{\pi}{4r_0} + \frac{2}{b} \right) }{1 +V_g^{-1}(0) (V_b -V_b(0) - Qt)}.
	\end{align}
	At the outlet of the Hele--Shaw cell (around the rim), we take the pressure of the liquid to be atmospheric, $p(r=R_c)=p_\atm$, such that the gauge liquid pressure $\Delta p=p-p_\atm$ is
	\begin{align}
		\Delta p &= 0, \qquad  |\boldsymbol{x}| \in R_c \label{eq:CellEdge}.
	\end{align}
	The flow is driven by the gauge liquid pressure difference between the liquid-gas interface and the outlet.
	
\end{subequations}

We non-dimensionalize our problem via
\begin{align}
	\vec{\hat{x}} = \frac{\vec{x}}{R_c}, \quad \hat{t} = \frac{Q}{\pi R_c^2 b} t, \quad \hat{p} = \frac{\pi b^3}{6 \mu Q} p, \quad \hat{\vec{v}} = \frac{2\pi R_c b}{Q} \vec{v}, \quad \hat{V} = \frac{V}{\pi R_c^2 b}, \label{eq:NonDimen}
\end{align}
leading to
\begin{subequations}
	\begin{alignat}{3}
		\hat{\nabla}^2 \hat{p} &= 0, &\hat{\boldsymbol{x}} &\in \mathbb{R}^2 \backslash \Omega(\hat{t}), \label{eq:Compressible1}\\
		\hat{v}_n &= - \frac{1}{1 - f_1} \hat{\grad} \hat{p} \cdot \vec{n},       &\hat{\boldsymbol{x}} &\in \partial \Omega(\hat{t}), \label{eq:Compressible2}\\
		\Delta\hat{p} &= \Delta\hat{p}_g(t) - \Ca^{-1} \left( \frac{\pi}{4} \hat{\kappa} + 2 \alpha f_2 \right), \qquad \qquad  &\hat{\boldsymbol{x}} &\in \partial \Omega(\hat{t}), \label{eq:Compressible3} \\
		\Delta\hat{p} &= 0, \qquad \qquad  &|\hat{\boldsymbol{x}}| &\in 1,  \label{eq:Compressible4}
	\end{alignat}
	where
	\begin{align}
		\Delta\hat{p}_g &= \frac{ 2\Comp^{-1}\left[\hat{t}+\mathcal{R}^2-\hat{V}_b(\hat{t})\right] + \Ca^{-1} \left(\frac{\pi}{4\mathcal{R}}+2\alpha\right) }{1 - \mathcal{V}^{-1} \left[\hat{t}+\mathcal{R}^2-\hat{V}_{b}(\hat{t})\right]} .  \label{eq:Compressible5}
	\end{align}
	The thin-film factors $f_1$ and $f_2$ are still given by Eq.~\eqref{eq:films1}, which is dimensionless. The dimensionless control parameters are
	\begin{align}
		\Ca = \frac{12 \alpha^2 \mu Q}{2\pi R_c b \gamma}, \quad \Comp = \frac{12 \mu Q V_g(0)}{\pi^2 R_c^2 b^4 p_\atm}, \quad \mathcal{V} = \frac{V_g(0)}{\pi R_c^2 b},  \qquad \alpha = \frac{R_c}{b},\qquad \mathrm{and} \qquad \mathcal{R} = \frac{r_0}{R_c},
	\end{align}
\end{subequations}
where $\Ca$ is the capillary number, often referred to as the `modified' capillary number as it incorporates the role of the cell aspect ratio $\alpha$ \citep[\textit{e.g.},][]{Park1985}. The parameter $\Comp$, which we refer to as the compressibility number, is directly analogous to the compressibility number identified in Ref.~\citep{Cuttle2023a}. Physically, $\Comp$ may be considered the ratio of viscous and compressive pressure scales or, equivalently, the ratio of the rate of viscous depressurisation (drainage) to that of compressive pressurisation (compression).

For comparison, we also consider the case where the gas is incompressible. In this limit, Eq.~\eqref{eq:Compressible5} is replaced by the constraint that $\Delta p_g$ must be chosen so that the actual injection rate $\hat{Q}_b \equiv \mathrm{d}\hat{V}_b/\mathrm{d}\hat{t} = 1$ at all times and the parameters $\Comp$ and $\mathcal{V}$ can be eliminated. This constraint can be enforced by choosing $\Delta p_g$ such that
\begin{align}
	\int_{0}^{2 \pi} -\frac{\p \hat{p}}{\p \hat{r}}\bigg\rvert_{\hat{r}=1} \, \textrm{d} \theta = 2\pi, \label{eq:Compressible7}
\end{align}
which ensures that the total rate of outflow is unity.

\subsection{Axisymmetric model} \label{sec:RedModel}

To gain insight into the coupling between gas compression and liquid displacement, we consider the axisymmetric limit in which the interface remains circular with radius $r = R_0(t)$. Equations \eqref{eq:Compressible1}-\eqref{eq:Compressible5} then reduce to
\begin{align}
	\frac{\text{d} \hat{R}_0}{\text{d} \hat{t}} &= \frac{\Delta\hat{p}_g - \Ca^{-1}\left(\frac{\pi}{4\hat{R}_0} + 2\alpha f_2\right) }{2(1-f_1)\hat{R}_0 \ln\left(1/\hat{R}_0\right)}, \label{eq:RadialCompressible1} \\
	\Delta\hat{p}_g &= \frac{ 2\Comp^{-1}\left(\hat{t}+\mathcal{R}^2-\hat{R}_0^2\right) + \Ca^{-1} \left(\frac{\pi}{4\mathcal{R}}+2\alpha\right)}{1 - \mathcal{V}^{-1}\left(\hat{t}+\mathcal{R}^2-\hat{R}_0^2\right) }. \label{eq:RadialCompressible2}
\end{align}
We can simplify further by taking $\mathcal{V} \gg1$ and $2\Comp^{-1} \gg \Ca^{-1} [\pi/(4\mathcal{R})+2\alpha]$; the former limit corresponds to an initial volume of gas that is much larger than the volume of liquid in the cell (i.e., a large gas reservoir) while the latter corresponds to negligible capillary pressure relative to atmospheric pressure, which also implies $2\Comp^{-1} \gg \Ca^{-1} [\pi/(4\hat{R}_0)+2\alpha f_2]$ and thus eliminates $f_2$ from the model. Additionally ignoring the kinematic impact of thin films (\textit{i.e.}, assuming $f_1\approx{}0$), Eqs.~\eqref{eq:RadialCompressible1}-\eqref{eq:RadialCompressible2} reduce to the ordinary differential equation
\begin{align}
	\frac{\text{d} \hat{R}_0}{\text{d}\hat{t}} &= \frac{\Delta\hat{p}_g}{ 2\hat{R}_0 \ln \left(1/\hat{R}_0\right)}, \quad \textrm{with} \quad \Delta\hat{p}_g = \frac{2}{\Comp} \left(\hat{t} + \mathcal{R}^2 -  \hat{R}_0^2 \right). \label{eq:CompressibleODE}
\end{align}
The evolution of the interface then depends on a single dimensionless parameter, the compressibility number $\Comp$. When the gas is incompressible ($\Comp \to 0$), Eq.~\eqref{eq:CompressibleODE} degenerates to $\hat{R}_0(\hat{t}) = \left( \mathcal{R}^2 + \hat{t} \right)^{1/2}$ and $\Delta\hat{p}_g=\ln(1/\hat{R}_0)$, as expected.

Throughout the present work, we focus on injection via the steady compression of a gas reservoir, as may be imposed by a syringe pump, for consistency with our experiments. However, another common experimental implementation of gas injection is from a high-pressure source regulated by a needle valve. In Appendix \ref{app:cylinder}, we derive a mathematical model for this alternative approach and show that the two methods are identical under the assumptions of the axisymmetric model.

\subsection{Numerical scheme}\label{sec:NumSche}

We solve Eqs.~\eqref{eq:Compressible1}-\eqref{eq:Compressible5} using the numerical scheme proposed by \citet{Morrow2021}, which we briefly summarize here. The scheme is based on the level-set method \cite{Osher1988}, where we construct a level-set function $\phi$ such that $\phi < 0$ in the gas region and $\phi > 0$ in the liquid region. We evolve $\phi$ via the level-set equation
\begin{align}
	\frac{\p \phi}{\p \hat{t}} + F \left| \hat{\nabla} \phi \right|  = 0, \label{eq:LevelSetEqn}
\end{align}
where
\begin{align}
	F = - \hat{\grad} \hat{p} \cdot \vec{n}, \label{eq:SpeedFn}
\end{align}
and $\vec{n} = \hat{\grad} \phi / |\hat{\grad} \phi|$ is the unit (outward) normal. This choice of speed function $F$ satisfies the kinematic boundary condition on the interface [Eq.~\eqref{eq:Compressible2}] and gives a continuous expression for $F$ in the liquid region $\vec{x} \in \mathbb{R}^2 \backslash \Omega(\hat{t})$. We extend $F$ into the gas region by solving the biharmonic equation \citep{Moroney2017}. To solve Eq.~\eqref{eq:LevelSetEqn}, we use a second-order essentially non-oscillatory scheme for the spatial derivatives, and integrate in time using second order total-variation-diminishing Runge-Kutta with $\Delta t = \Delta x / [4 \max|F|]$. To maintain $\phi$ as a signed distance function such that $|\hat{\nabla} \phi| = 1$, we occasionally perform reinitialization~\citep{Morrow2021}. We solve Eq.~\eqref{eq:Compressible2} for the liquid pressure via a finite difference stencil. Following \citet{Gibou2002}, we modify the stencil at nodes adjacent to the gas--liquid interface to incorporate the dynamic boundary condition [Eq.\eqref{eq:Compressible3}], where the signed curvature of the interface is $\kappa = \grad \cdot \vec{n}$. Further, the volume of the gas region is computed with
\begin{align}
	\hat{V}_g = \int_{\mathbb{R}^2} (1-f) H(\phi) \, \textrm{d} \hat{V},
\end{align}
where
\begin{align}
	H = \begin{cases}
		0  & \textrm{ if } \phi < -\delta \\
		1 & \textrm{ if }  \phi > \delta \\
		\frac{1}{2} \left[ 1 + \frac{\phi}{\delta} + \frac{1}{\pi} \sin \left(\frac{\pi \phi}{\delta}\right) \right] & \text{ otherwise.} \\
	\end{cases},
\end{align}
Here, $\delta = 1.5 \Delta x$, and $f$ is the proportion of the Hele-Shaw cell filled with liquid at each node, as determined from $f_1$ [see Eq.~\eqref{eq:films1}]. When the gas is incompressible, we discretise the integral in Eq.~\eqref{eq:Compressible7} via the trapezoidal rule. We solve the resulting system of linear equations via LU decomposition. All simulations are performed with the initial condition
\begin{align}
	\hat{R}(\theta, 0) = \mathcal{R} \left\{ 1 + \sum_{n=2}^{12} \varepsilon_n \cos \left[ n (\theta - 2 \pi \theta_n) \right] \right\},  \label{eq:InitialCondition}
\end{align}
where $\theta_n$ and $\varepsilon_n$ are selected at random from uniform distributions on the intervals (0, 1) and ($5 \times 10^{-4}$, $10^{-3}$), respectively. Simulations are performed on the computational domain $0 \le r \le 1$ and $0 \le \theta < 2 \pi$ using $1000 \times 1000$ equally spaced nodes. Simulations are stopped when the maximum radius of the interface is 0.99; we denote the time at which this occurs as $\hat{t} = \hat{t}_f$.

\section{Experimental methods}\label{sec:Exp}

\subsection{Set-up}\label{sec:Setup}

Experiments are performed in a Hele-Shaw cell comprising two glass plates of radius $R_c=105$~mm. We impose a gap $b=0.42\pm0.01$~mm between the plates using a plastic spacer. The spacer supports the outer 5~mm of the plates and, in doing so, obstructs a small fraction of the outflow area. Note that this partial obstruction is not included in the simulations, and is believed to contribute to the systematically greater pressures recorded in experiments, although this discrepancy could instead result from qualitative differences in fingering patterns (see \S~\ref{sec:BulkQP} and \S~\ref{sec:Nonlinear}).

The cell is initially filled with 1,000~cSt silicone oil (dynamic viscosity $\mu=0.97$~Pa~s and surface tension $\gamma=21$~mN~m$^{-1}$ at laboratory temperature $22\pm1^{\circ}$C; Sigma), which is dyed (Sudan III; Merck) and filtered. We impose a fixed hydrostatic pressure at the outer rim of the cell (the outlet) by surrounding the cell with a shallow oil reservoir, filled to a depth approximately 1~mm above the top of the gap, which is maintained using an overflow.

The cell is connected to an air reservoir via a 2~mm diameter injection port. The reservoir comprises two 50~mL airtight glass syringes (1050TTL; Hamilton), along with stiff connective tubing of internal volume $12\pm1$~mL (Legris). Short sections of connective tubing (Tygon) are used sparingly to minimise pressure-induced changes in the volume of the air reservoir itself. The total initial volume of the air reservoir (tubing and syringes) is set to $V_{\res}(0)\in\{25,50,100,200\}$~mL, with a relative error of 2-4\%. To achieve $V_{\res}(0)$=200~mL, we connect an additional sealed acrylic box of internal volume $120\pm3$~mL. The total internal volume of the box and tubing was measured via changes in air pressure during controlled compression tests with the system closed. These tests also suggest that air leakage was negligible over the timescales and pressures of our experiments.

Prior to each experiment, we introduced a circular precursor bubble of initial radius $R_0(0)=2.7\pm0.1$~mm by injecting air very slowly, such that no significant pre-compression of the air was introduced. The initial air pressure is taken to be atmospheric, neglecting the small hydrostatic pressure imposed by the oil reservoir and the Laplace pressure jump at the interface. The initial volume of the bubble, around 0.01~mL, is negligible compared with $V_{\res}(0)$.

To conduct an experiment, the air was then compressed using a syringe pump (AL-4000; WPI) at a fixed volumetric rate $Q\in\{1.25,2.50,5.00,10.0\}$~mL~min$^{-1}$. The gauge pressure $\Delta p_g$ of the air relative to atmospheric pressure $p_\atm$ was recorded using a pressure sensor (0-15~PSI; Honeywell) via a USB DAQ (U6; LabJack) at a frequency of approximately 20~Hz. The pressurised air drove oil out of the cell and into the surrounding oil reservoir.

We imaged the motion of the interface using a CMOS camera (acA4096-30um; Basler) mounted vertically below the cell and recording at a fixed frame rate of 1.5--12~frames per second (fps), depending on $Q$, and at a spatial resolution 91~$\mu$m~per~pixel. Over each experiment, we typically recorded 400~frames. The cell was backlit using a custom array of LEDs (Wholesale LED Lights), diffused through opalescent acrylic (Sheet Plastics) and a blue filter (Stage Depot) to enhance contrast with the dyed oil. Ambient light from the laboratory was blocked by cloaking the set-up in opaque fabric (BK5; Thorlabs).

Each experiment was performed twice to assess reproducibility. We found that bulk displacement measurements, such as the evolution of the injection rate and the air pressure, were highly reproducible between experiments despite significant variation in the fingering patterns and their associated metrics after onset (see \S~\ref{sec:Nonlinear} and Appendix~\ref{app:reprod}).

\subsection{Data processing}\label{sec:DataProc}

The volume of the air in the cell $V_b$ was calculated from the air pressure. We did so by modelling the air as a fixed mass of isothermal ideal gas, such that $[p_\atm+\Delta p_g(t)]V_g(t)=p_\atm V_\res(0)$. Taking $V_g(0) = V_{\textrm{res}}(0)$ and $p_g(0) = p_{\textrm{atm}}$ by neglecting the initial volume $V_b(0)$ of the bubble and the initial Laplace and hydrostatic pressures, respectively, introduces negligible error in this calculation. The total volume of air, which changes due to both compression of the reservoir and invasion into the flow cell is $V_g(t)=V_{\res}(0)-Qt+V_b(t)-V_b(0)$, so that
\begin{align}\label{eq:VolBub}
	V_b(t)=V_b(0) + Qt-V_{\res}(0)\left[1-\left(\frac{p_\atm}{p_\atm +\Delta p_g(t)}\right)\right].
\end{align}
Accounting for $V_b(0)$ at this step ensures that the initial value of $V_b(t)$ is accurate. In Appendix~\ref{app:adiabatic} we compare Eq.~\eqref{eq:VolBub} with an adiabatic model to show that the assumption of isothermal compression is justified for our experiments. The actual time-dependent injection rate of air into the cell $Q_b=\mathrm{d}V_b/\mathrm{d}t$ (which is equivalent to the flow rate of liquid out of the cell) was calculated at each recorded $V_b(t)$ data point by taking a second-order polynomial least-squares fit to the data on either side of that point. The first derivative of the fitted function was then taken as the local value of $Q_b(t)$; the size of the fitting window was automatically adjusted to minimise a $\chi^2$ fitting parameter that avoided over- or underfitting to the data.

\begin{figure}
	\centering
	\includegraphics[width=0.7\linewidth]{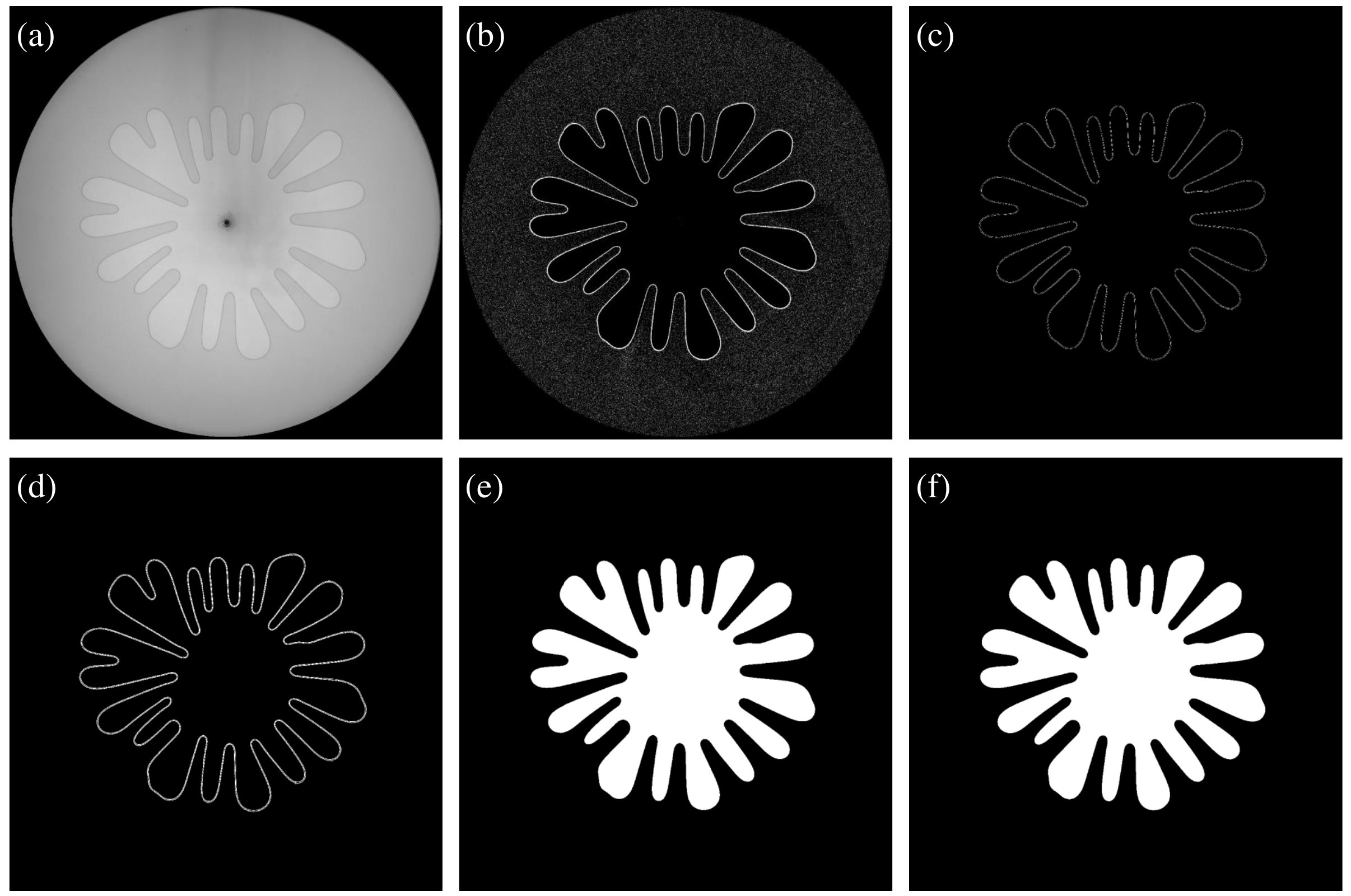}
	\caption{Key stages of image analysis: (a)~Raw image, (b)~background subtraction and contrast enhancement, (c)~edge detection, (d)~dilation, (e)~filling, and (f)~erosion. This image is taken from an experiment with $Q=5$~mL/min and $V_\res(0)=200$~mL at $t=83.3$~s. \label{fig:Imaging} }
\end{figure}

Recorded experimental images were processed in \texttt{MATLAB} R2020b to recover the two-dimensional area of the gas region and the shape of the interface as functions of time. The key steps of the algorithm are illustrated in Fig~\ref{fig:Imaging}. From each raw image [Fig~\ref{fig:Imaging}(a)], a reference image of the cell prior to air injection was subtracted. The resulting difference image was then contrast-enhanced using \texttt{imadjust} to facilitate the isolation of the interface [Fig~\ref{fig:Imaging}(b)]. Pixels along the interface were identified using the \texttt{edge} function with a \texttt{canny} filter [Fig~\ref{fig:Imaging}(c)]. The resulting binary image was dilated by applying \texttt{imerode} with a circular structuring element of radius one pixel to a negative of the edge-detected image, which ensured connectivity of the interface [Fig~\ref{fig:Imaging}(d)]. A binary image of the air region was generated by filling the connected interface contour using \texttt{bwconncomp} [Fig~\ref{fig:Imaging}(e)] and then eroding using \texttt{imerode} with the same structure element to compensate for the earlier dilation step [Fig~\ref{fig:Imaging}(f)]. The area and perimeter of the air region were then calculated using \texttt{regionprops} and converted to dimensional units according to the spatial resolution of the camera. We process images up to the first frame when the maximum radial coordinate of the interface $\max(R)$ is greater than $0.9R_c$, beyond which interaction between the interface and the spacer became visually noticeable. We refer to the moment when this occurs as the near-breakout time $t_{0.9}$.

\section{Dimensional and non-dimensional parameter ranges}

Going forward, we discuss results primarily in terms of $\Ca$ and $\Comp$. In both experiments and simulations, however, we choose to vary the dimensional nominal injection rate $Q$ and initial gas volume $V_g(0)$ for practical reasons. Moreover, while $\Ca$ depends only on $Q$ and not on $V_g(0)$, $\Comp$ is proportional to the product $QV_g(0)$. To vary $\Comp\propto Q V_g(0)$ while keeping $\Ca\propto Q$ fixed, we fix $Q$ and vary the initial gas volume within the range $V_g(0)\in$\{3.125, 6.25, 12.5, \textbf{25, 50, 100, 200}, 400, 800, 1600\}~mL, equivalent to $\mathcal{V}\in$\{0.215, 0.430, 0.859, \textbf{1.72, 3.44, 6.87, 13.7}, 27.5, 55, 110\} (experiments were only performed at bold values). To vary $\Ca$ while keeping $\Comp$ fixed, we vary $Q$ while fixing the product $QV_g(0)$. We restrict our simulations to the experimental values of $Q$, corresponding to $\Ca\in\{2.61\times10^3$, $5.21\times10^3$, $1.04\times10^4$, $2.08\times10^4$\}. Values of $\mathcal{V}$ are given in the caption of each figure, where appropriate. The remaining parameters are fixed at $\alpha=250$ and $\mathcal{R}=0.025$, with dimensional values listed in \S~\ref{sec:Setup}.

\section{Results}\label{sec:Results}

Section~\ref{sec:Results} is organised as follows. We begin in \S~\ref{sec:GenObs} by describing the key observations of compression-driven viscous fingering. In \S~\ref{sec:BulkQP}, we then consider the `bulk' displacement dynamics, and specifically the unsteady injection rate and injection pressure, that arise from compression-driven displacement in a radial Hele-Shaw cell. We show that the axisymmetric model derived in \S~\ref{sec:RedModel} qualitatively captures the variation in bulk displacement dynamics with increasing $\Comp$. Comparison between experiments, simulations and the axisymmetric model demonstrates that the bulk displacement dynamics of the full system are also controlled primarily by $\Comp$ and are remarkably insensitive to both variations in $\Ca$ and the presence of viscous fingering. In \S~\ref{sec:Regimes}, we examine the underlying dynamical-systems structure that dictates the unsteady injection rate in the axisymmetric model. The dynamical regimes described in this framework qualitatively predict the bursts of high flux observed at $\Comp\gtrsim1$ both in experiments and simulations. In \S~\ref{sec:Stability}, we perform a linear stability analysis. At sufficiently large $\Ca$, our analysis suggests that compressibility significantly delays the onset of viscous fingering relative to an incompressible system; we also identify a weaker effect at low $\Ca$, where compressibility may instead promote onset to comparatively smaller radii. In \S~\ref{sec:Nonlinear}, we examine the growth of viscous fingers in both experiments and simulations. We observe a strong delay in the onset with increasing $\Comp$, consistent with the predictions of our linear stability analysis. Finally, we discuss the qualitative impacts of compressibility on the nonlinear evolution of the fingering pattern in both experiments and simulations.

\subsection{Features of compression-driven viscous fingering} \label{sec:GenObs}

\begin{figure}
	\centering
	\includegraphics[width=0.9\linewidth]{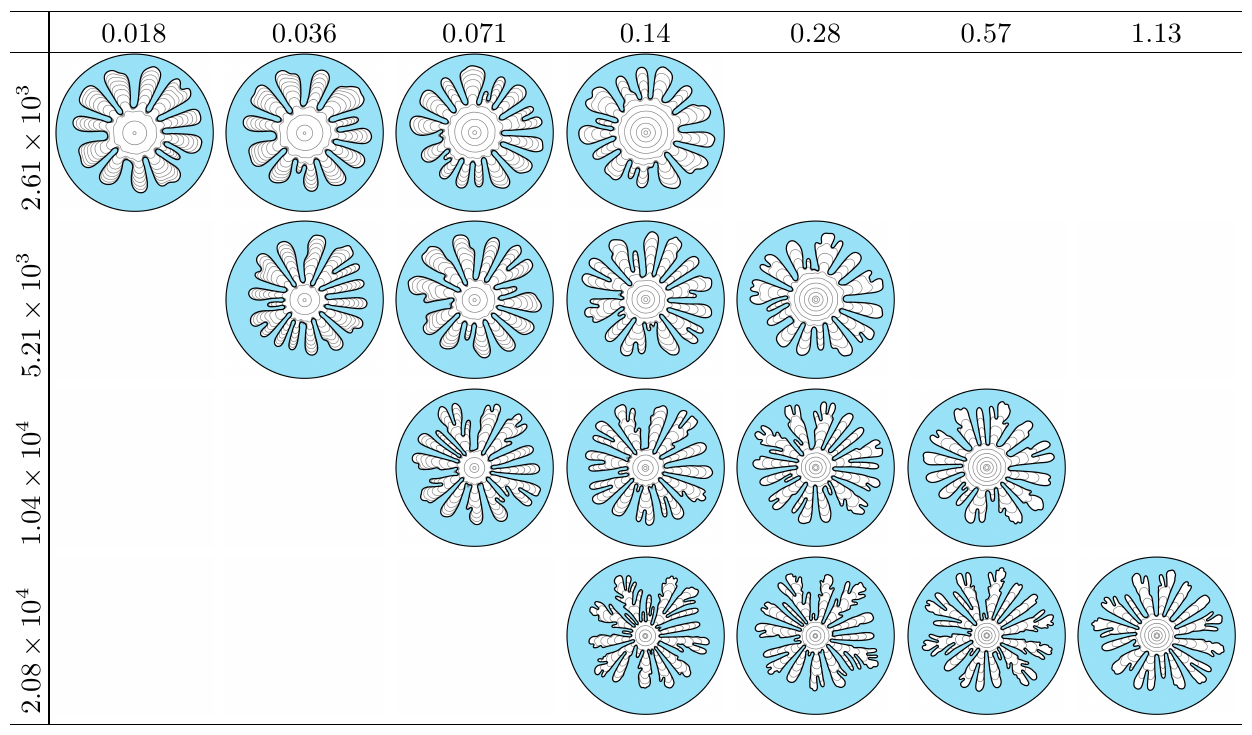}
	\caption{Experimental fingering patterns for $\Ca$ increasing top to bottom and $\Comp$  increasing left to right. The corresponding dimensional parameters are (top to bottom) $Q = [1.25, 2.5, 5, 10]$~mL/min and (along each north-west/south-east diagonal) $V_\res(0) = [25, 50, 100, 200]$~mL, with $R_c = 10.5$~cm, $\mu=0.97$~Pa~s, $\gamma=21$~mN~m$^{-1}$, and $b = 0.42$~mm. Profiles are plotted in equal time intervals of $t_{0.9}/10$.}
	\label{fig:Experiments}
\end{figure}

\begin{figure}
	\centering
	\includegraphics[width=1.0\linewidth]{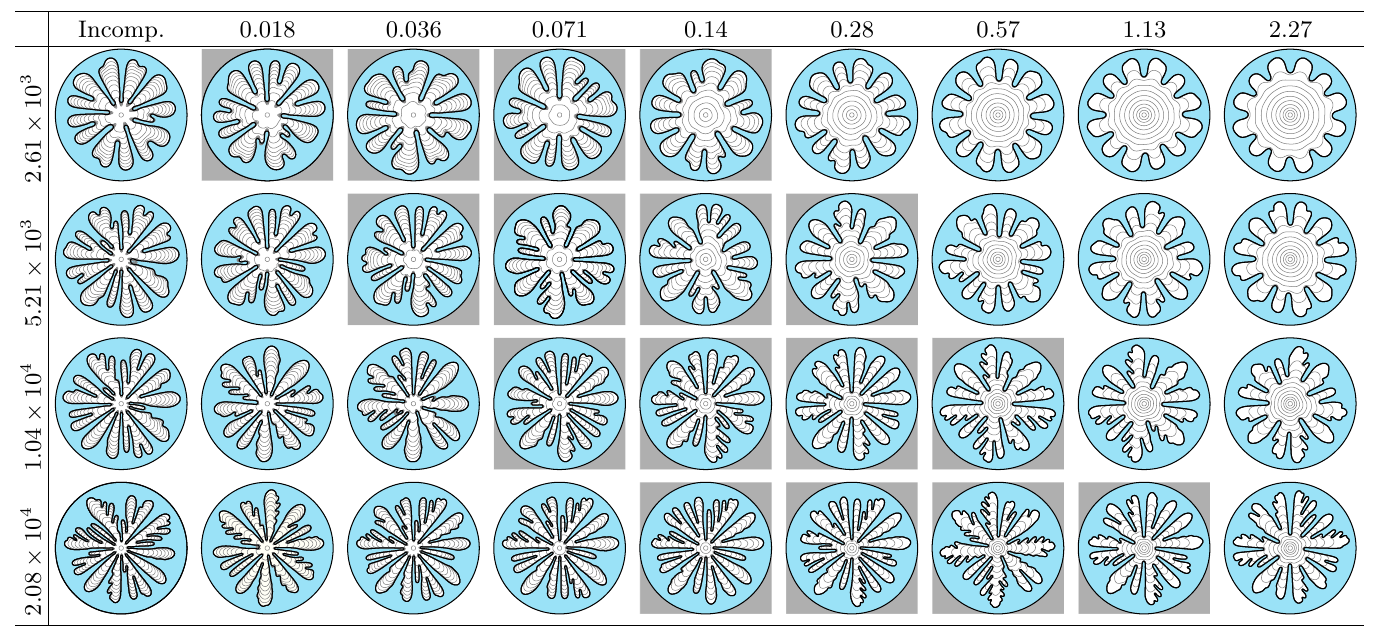}
	\caption{Example numerical solutions for $\Ca$ increasing top to bottom. Column one shows solutions of the incompressible model, while columns two to eight show numerical solutions of the compressible model Eqs.~\eqref{eq:Compressible1}-\eqref{eq:Compressible5} with $\Comp$ increasing left to right. Profiles are plotted in equal time intervals of $t_{0.9}/10$.  Simulations with shaded background correspond to the same parameter values as the experiments shown in Fig.~\ref{fig:Experiments}.}
	\label{fig:Simulations}
\end{figure}

We begin by briefly examining the qualitative impact of compressibility on viscous fingering. In Figs.~\ref{fig:Experiments} and \ref{fig:Simulations}, we show the evolution of the fingering pattern across all of our experiments and simulations, respectively. For the experiments [Fig.~\ref{fig:Experiments}], each image includes the observed interface at time intervals of $t_{0.9}/10$, where $t_{0.9}$ is the near-breakout time at which the experiment is concluded (\S~\ref{sec:DataProc}). Rows correspond to fixed $\Ca$, while columns correspond to fixed compressibility number. In terms of dimensional quantities, we can interpret this arrangement as each row having a fixed nominal injection rate $Q$ with the initial gas reservoir size fixed along each north-west/south-east diagonal and increasing from left to right in each row. The behaviour for increasing $\Ca$ (top to bottom) is as expected from classical work on viscous fingering \citep{Chen1987}: The fingering pattern becomes more severe as $\Ca$ increases in the sense that we observe more and narrower fingers, as well as increasing instances of tip-splitting and side branching; Additionally, the onset of fingering (i.e., the point at which the interface deviates noticeably from a circle) appears to occur at smaller radii for larger $\Ca$.

It is striking, however, that we see a similar decrease in onset radius with decreasing $\Comp$ at fixed $\Ca$ (i.e., decreasing the air reservoir volume while fixing the nominal injection rate), which corresponds to moving from right to left along a given row. In other words, increasing $\Comp$ at fixed $\Ca$, in this case by using a larger air reservoir, appears to systematically delay the onset of viscous fingering. Varying the initial reservoir volume $V_\res(0)$ does not change $\Ca$, so its impact on viscous fingering is not considered in classical studies. Rather, changing $V_{\textrm{res}}(0)$ changes how `compressible' the system is~\citep{Sandnes2011}, as measured by the value of $\mathcal{C}$. This change in compressibility modifies the actual time-dependent injection rate (as distinct from the nominal injection rate $Q$) due to coupling between viscous forces in the displaced liquid and compressive forces in the air \citep{Cuttle2023a}. Our focus below is to formally rationalise and quantify the impact of compressibility in this system.

In addition to our experimental results, we conducted an extensive set of simulations over a wider range of parameters. These simulation results are shown in Figure~\ref{fig:Simulations}, again with rows and columns ordered by $\Ca$ and $\Comp$, respectively. The first column shows the results for an incompressible system. Images with a shaded square background indicate simulations conducted at the same parameters as the experiments in Fig.~\ref{fig:Experiments}. Our simulations and experiments are in strong qualitative agreement, showing consistent variations in patterns as $\Ca$ and $\Comp$ are varied. We analyse and compare these results quantitatively in \S~\ref{sec:Nonlinear}.

\subsection{Bulk displacement dynamics} \label{sec:BulkQP}

\begin{figure}
	\centering
	\includegraphics[width=0.75\linewidth]{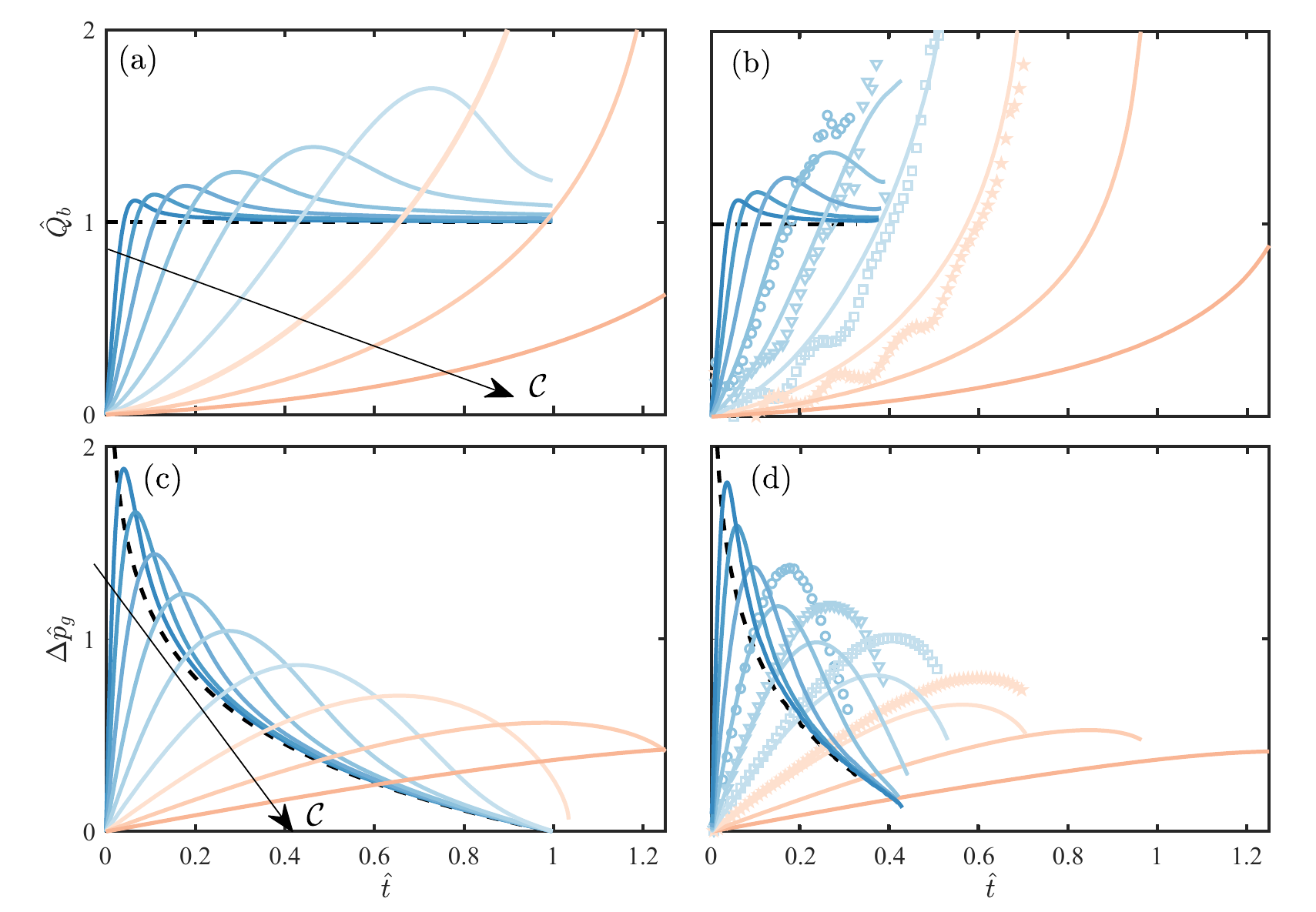}
	\caption{Compressibility number $\Comp$ has a strong impact on the evolution of injection rate $\hat{Q}_b$ and gas gauge pressure $\Delta\hat{p}_g$. Here, we plot (a) $\hat{Q}_b$ and (c) $\Delta\hat{p}_g$ as functions of time $\hat{t}$ from from the numerical solution to the axisymmetric model [Eq.~\eqref{eq:CompressibleODE}] with $\mathcal{R} = 0.025$ and (blue to red) $\mathcal{C} = 0.018$, $0.036$, $0.071$, $0.14$, $0.28$, $0.57$, $1.13$, $2.27$, and $4.54$. The incompressible solution is shown for comparison (dashed black curves). Panels (b) and (d) show the same quantities from numerical simulations (solid curves) and experiments (symbols) with viscous fingering at $\Ca = 2.08 \times 10^4$  ($Q=10$~mL/min and $V_\res(0)\in\{25,50,100,200\}$~mL; row 4 of Figs.~\ref{fig:Experiments} and \ref{fig:Simulations}). Blue and red shades indicate $\Comp<1$ and $\Comp>1$, respectively.}
	\label{fig:VolPressureComp}
\end{figure}

The dynamics of the axisymmetric model [Eq.~\eqref{eq:CompressibleODE}] are illustrated in Fig.~\ref{fig:VolPressureComp}, which shows the evolution of the nondimensional injection rate $\hat{Q}_b$ [Fig.~\ref{fig:VolPressureComp}(a)] and the gauge gas pressure $\Delta\hat{p}_g$ [Fig.~\ref{fig:VolPressureComp}(c)] for different compressibility numbers $\Comp$. For reference, the incompressible solution for a circular interface is also shown (dashed black curves), for which $\hat{Q}_b=1$ and $\Delta\hat{p}_g$ monotonically decreases as liquid drains and viscous resistance decreases. In the axisymmetric model, by contrast, the injection rate varies strongly as the interface advances and the gas pressure evolves non-monotonically. This behaviour is due to the basic coupling between viscous displacement and compressive pressurisation (fingering is absent from this model). Initially, the gas compresses and pressurises, such that $\hat{Q}_b$ and $\Delta\hat{p}_g$ increase gradually from initial values of zero. As the interface advances and drives liquid out, the resistance to flow decreases. The injection rate eventually exceeds the nominal injection rate ($\hat{Q}_b>1$), at which point the gas begins to expand and $\Delta\hat{p}_g$ decreases. For $\Comp\ll1$, the compressible dynamics differ only weakly from the incompressible dynamics, with $\hat{Q}_b$ rapidly reaching and then exceeding the nominal flux, before relaxing back toward the incompressible solution $\hat{Q}_b=1$. As $\Comp$ approaches unity, the maximum in $\hat{Q}_b$ increases and occurs at later times. For $\Comp>1$, the maximum in $\hat{Q}_b$ vanishes, and the injection rate instead increases monotonically, diverging as the interface escapes the cell at the moment of breakout (when $\hat{R}_0=1$). The qualitative change in dynamics around $\Comp=1$ is consistent with the distinct dynamical regimes observed for compression-driven displacement in a capillary tube~\citep{Cuttle2023a}, as discussed further in \S~\ref{sec:Regimes}.

We show in Fig.~\ref{fig:VolPressureComp}(b, d) that the full numerical simulations (solid lines) and the experiments (symbols) with viscous fingering both exhibit qualitatively similar displacement dynamics in $\hat{Q}_b$ and $\Delta\hat{p}_g$. The presence of fingers in the experiments and simulations leads to an earlier breakout than in the axisymmetric model. In addition, the transition from non-monotonic to monotonically increasing $\hat{Q}_b$ occurs at a slightly lower value of $\Comp$ in the simulations and experiments than in the axisymmetric model. Otherwise, the evolution of $Q_b$ and $\Delta\hat{p}_g$ and the variation with $\Comp$ are strikingly similar. This agreement suggests that viscous fingering has only a weak influence on the underlying displacement dynamics, primarily leading to earlier breakout compared with the axisymmetric model [Eq.~\eqref{eq:CompressibleODE}]. Furthermore, the quantitative agreement between experiments and corresponding simulations is in contrast with the visually distinct fingering patterns generated in each case; comparing the bottom rows of Figs.~\ref{fig:Experiments} and \ref{fig:Simulations}, we see that experiments produce fingering patterns with greater instances of tip-splitting and side-branching, resulting in more severely distorted interfaces (see \S~\ref{sec:Nonlinear}).

\begin{figure}
	\centering
	\includegraphics[width=0.8\linewidth]{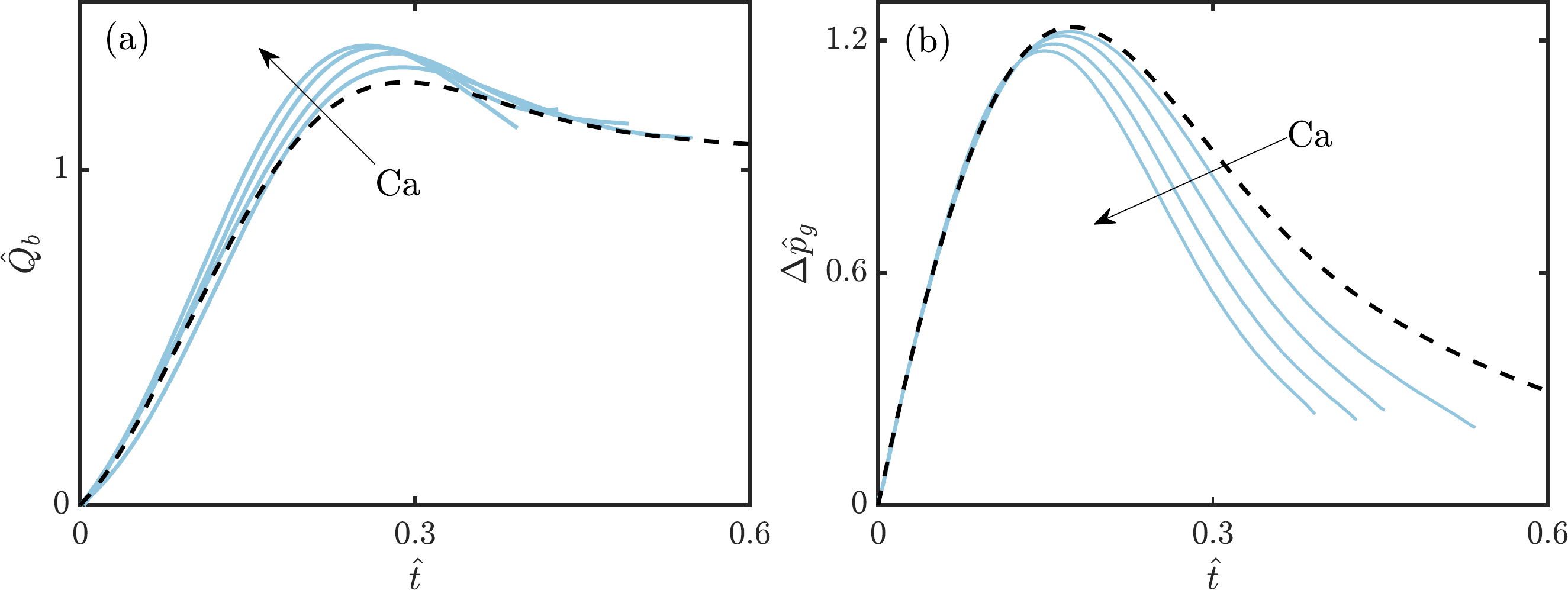}
	\caption{Capillary number $\Ca$ has a modest impact on the evolution of injection rate $\hat{Q}_b$ and gauge gas pressure $\Delta\hat{p}_g$. Here, we plot (a) $\hat{Q}_b$ and (b) $\Delta\hat{p}_g$ for $\mathcal{C} = 0.14$ from numerical simulations of viscous fingering at $\Ca\in\{2.61\times10^3$, $5.21\times10^3$, $1.04\times10^4$, $2.08\times10^4$\} ($\mathcal{V}\in\{1.72,3.44,6.87,13.7\}$; column 5 in Fig.~\ref{fig:Simulations}) and from the axisymmetric model (dashed black).}
	\label{fig:VolPressureCa}
\end{figure}

As shown in Fig.~\ref{fig:VolPressureCa}, the evolution of $\hat{Q}_b$ [Fig.~\ref{fig:VolPressureCa}(a)] and $\Delta\hat{p}_g$ [Fig.~\ref{fig:VolPressureCa}(b)] at fixed $\Comp$ is also only weakly modified by varying $\Ca$ over an order of magnitude, despite the strong variation in fingering patterns (see fourth column in Fig.~\ref{fig:Experiments} and fifth column of  Fig.~\ref{fig:Simulations}). The dominant effect of fingering on compression-driven displacement is to induce an increasingly early breakout as $\Ca$ increases. Larger values of $\Ca$ also lead to systematically lower values of $\Delta\hat{p}_g$, consistent with the fact that more severe fingering patterns bypass an increasingly large fraction of the liquid. Hence, the bulk displacement dynamics, in terms of injection rates and driving pressures, are remarkably insensitive to viscous fingering, and are governed primarily by $\Comp$. As noted in \S~\ref{sec:Exp}, the systematically greater pressure recorded in experiments [Fig.~\ref{fig:VolPressureComp}(d)] may derive in part from the geometry of the spacer used to impose the gap, which was not accounted for in the simulations. However, differences in the fingering patterns [Figs.~\ref{fig:Experiments} and \ref{fig:Simulations}] may also influence the pressure evolution, as illustrated by the results of Fig.~\ref{fig:VolPressureCa}(b) (see \S~\ref{sec:Nonlinear}).

\subsection{Dynamical systems framework}\label{sec:Regimes}

The axisymmetric model exhibits two dynamical regimes, characterised by the change from non-monotonic to monotonically increasing $\hat{Q}_b$ around $\Comp\approx1$ [Fig.~\ref{fig:VolPressureComp}$(a)$]. We next rationalise these regimes in terms of a general dynamical systems framework. A similar treatment was originally applied to compression-driven displacement in a capillary tube by~\citet{Cuttle2023a}; we take the same approach here to describe the dynamics of axisymmetric displacement in a Hele-Shaw cell, as embodied by the axisymmetric model [Eq.~\eqref{eq:CompressibleODE}].

We start by considering the injection rate, which for a circular interface is $\hat{Q}_b=2\hat{R}_0(\mathrm{d}\hat{R}_0/\mathrm{d}\hat{t})$. From the axisymmetric model, $\hat{Q}_b=\Delta\hat{p}_g/\hat{\omega}$, where we have introduced the resistance $\hat{\omega}=\ln\left(1/\hat{R}_0\right)$ by analogy with Ohm's law. We can then write the axisymmetric model as
\begin{align}\label{eq:RelRates}
	\RRC{\hat{Q}_b} = \RRC{\Delta\hat{p}_g} - \RRC{\hat{\omega}}=\frac{2}{\hat{\omega}\Comp}\left(\frac{1}{\hat{Q}_b}-1+\frac{\Comp\hat{Q}_b}{4\hat{R}_0^2}\right),
\end{align}
where $\RRC{\hat{x}}=\dot{x}/\hat{x}$ is the relative rate of change of the variable $\hat{x}$, with $\dot{x}=\mathrm{d}\hat{x}/\mathrm{d}\hat{t}$. The axisymmetric model admits two trivial solutions, for which the driving compressive force and the opposing viscous resistance decrease at the same relative rate [i.e., $\lambda(\hat{Q}_b)=0$]. These are
\begin{align}\label{eq:TrivSol}
	\bar{Q}_{\pm}=2\hat{R}_0\left(\frac{\hat{R}_0\pm\sqrt{\hat{R}_0^2-\Comp}}{\Comp}\right),
\end{align}
which satisfy $\RRC{\bar{Q}_\pm}=0$.

\begin{figure}
	\centering
	\includegraphics[width=0.8\linewidth]{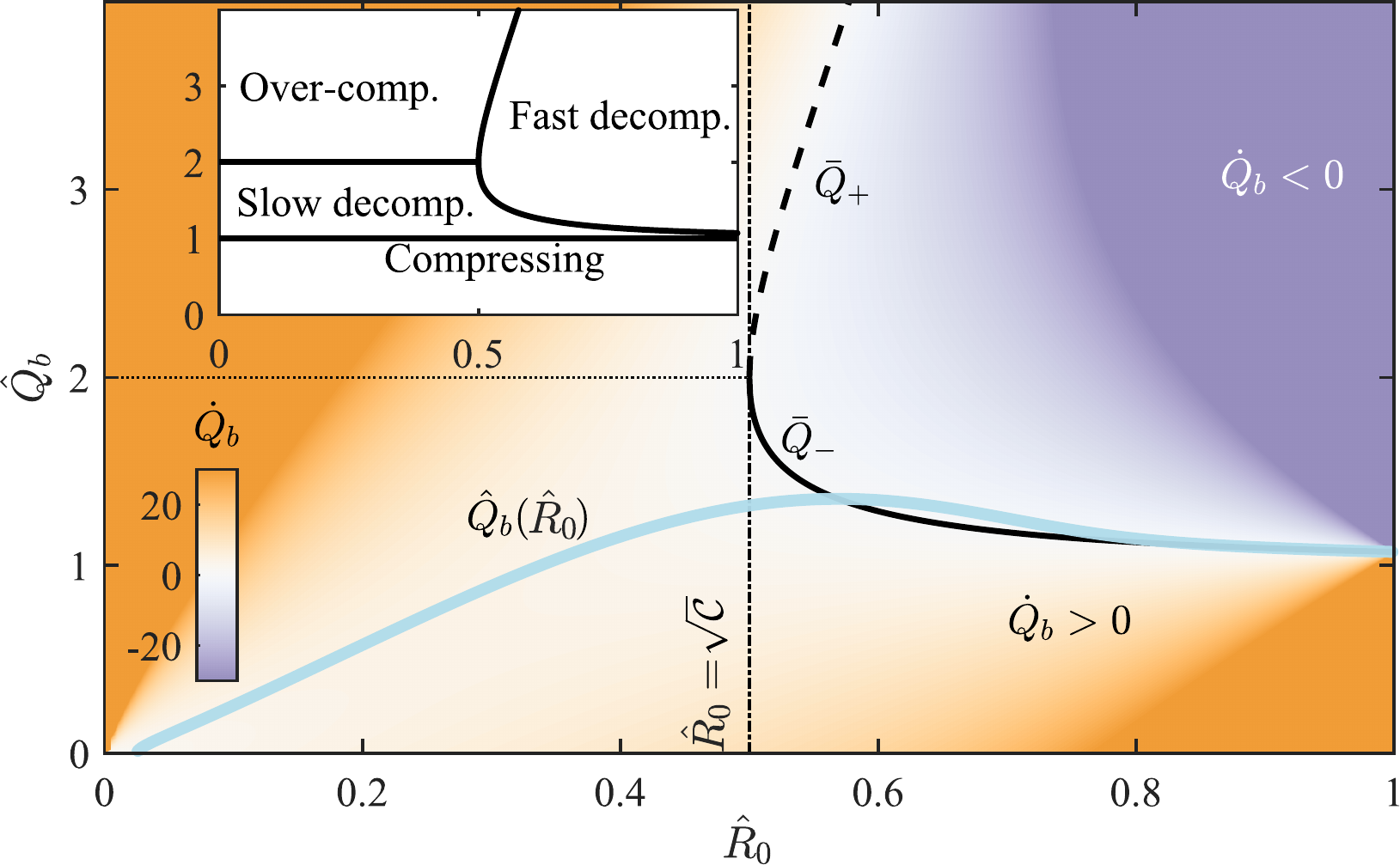}
	\caption{Phase-space representation of the axisymmetric model [Eq.~\eqref{eq:CompressibleODE}], in terms of injection rate $\hat{Q}_b$ and interface radius $\hat{R}_0$, for $\Comp=0.25$. The local time derivative of injection rate $\dot{Q}_b$ is indicated by the colormap (see colorbar). Stable and unstable trivial solutions $\bar{Q}_\pm$ are shown as solid and dashed black lines, respectively. The $\hat{R}_0$ and $\hat{Q}_b$ coordinates of the saddle-node bifurcation point are indicated by dot-dashed and dotted lines, respectively. A solution of the axisymmetric model for $\Comp=0.25$ and $\mathcal{R}=0.01$ is plotted as a thick blue curve. Regions of the phase space for which $\dot{Q}_b$ is positive or negative are shaded orange or purple, respectively (see colorbar). When $\RRC{\Delta\hat{p}_g}>\RRC{\hat{\omega}}$, the interface accelerates, $\dot{Q}_b>0$. This occurs in three distinct subregions (see inset): when $\hat{Q}_b<1$ because the gas is compressing while resistance decreases ($\Delta\dot{p}_g>0$ and $\dot{\omega}<0$); when $1<\hat{Q}_b<\bar{Q}_-$ because the gas is decompressing slowly ($\Delta\dot{p}_g<0$ is small); and when $\hat{Q}_b>\bar{Q}_+$ because the gas is over-compressed ($\Delta\hat{p}_g$ is large). When $\RRC{\Delta\hat{p}_g}<\RRC{\hat{\omega}}$, the interface decelerates, $\dot{Q}_b<0$. This occurs only in the region $\bar{Q}_-<\hat{Q}_b<\bar{Q}_+$, where the gas decompresses quickly ($\Delta\dot{p}_g<0$ and $|\Delta\dot{p}_g|$ is large).}
	\label{fig:PhaseSpace}
\end{figure}

The influence of these trivial solutions on compressible displacement dynamics is illustrated by the phase-space plot in Fig.~\ref{fig:PhaseSpace}, where $\bar{Q}_\pm(\hat{R}_0)$ are plotted as dashed and solid black curves, respectively, for $\Comp=0.25$. The trivial solutions separate regions where $\dot{Q}_b$ is positive (orange) or negative (purple). The local change of sign in $\dot{Q}_b$ determines the stability of each trivial solution. Specifically, $\bar{Q}_+$ is a repeller and $\bar{Q}_-$ is an attractor, in the sense that small perturbations grow or decay, respectively. A solution of the axisymmetric model is plotted as a faint blue curve, displaying the characteristic nonmonotonic variation in $\hat{Q}_b$ with $\hat{R}_0$ and showing that, rather than tending back toward the incompressible solution ($\hat{Q}_b=1$), the system is drawn onto the attractive solution $\bar{Q}_-$. 

Due to their $\hat{R}_0$-dependence, the trivial solutions only exist for $\hat{R}_0\ge\sqrt{\Comp}$. At $\hat{R}_0=\sqrt{\Comp}$, the two branches $\bar{Q}_\pm$ meet and annihilate at a saddle-node bifurcation. For $\hat{R}_0<\sqrt{\Comp}$ there are no trivial solutions: $\dot{Q}_b>0$ for all $\hat{Q}_b$, so that the interface accelerates monotonically. Because $\hat{R}_0$ must be less than unity, $\Comp>1$ implies that there is no attractive solution and the interface accelerates monotonically toward breakout. Even for $\Comp<1$, however, the dynamics may fail to converge onto $\bar{Q}_-$, depending on the initial radius $\mathcal{R}$. Computing the basin of attraction in terms of $\mathcal{R}$ for a given $\Comp$ is only possible numerically, and is beyond the scope of this study. A brief exploration suggests that, for $\mathcal{R}=0.01$, the system fails to converge onto $\bar{Q}_-$ when $\Comp\ge0.92$. Therefore, $\Comp=1$ is an upper bound on the critical value at which the stability of the flow changes, unlike in a capillary tube where the critical value $\Comp=1$ is a precise indicator of the flow regime~\citep{Cuttle2023a}.

The stable trajectories that do converge onto $\bar{Q}_-$ terminate with a breakout flux of $\hat{Q}_b(\hat{R}_0=1)\equiv2\left(1-\sqrt{1-\Comp}\right)/\Comp$ because the resistance $\hat{\omega}$ vanishes at the moment of breakout, driving the system exactly onto $\bar{Q}_-$. Similarly, for unstable trajectories that fail to converge onto $\bar{Q}_-$, the vanishing resistance drives divergent $\RRC{\hat{Q}_b}$, and hence divergent $\hat{Q}_b$, in the absence of a local attractive solution.

\begin{figure}
	\centering
	\includegraphics[width=0.80\linewidth]{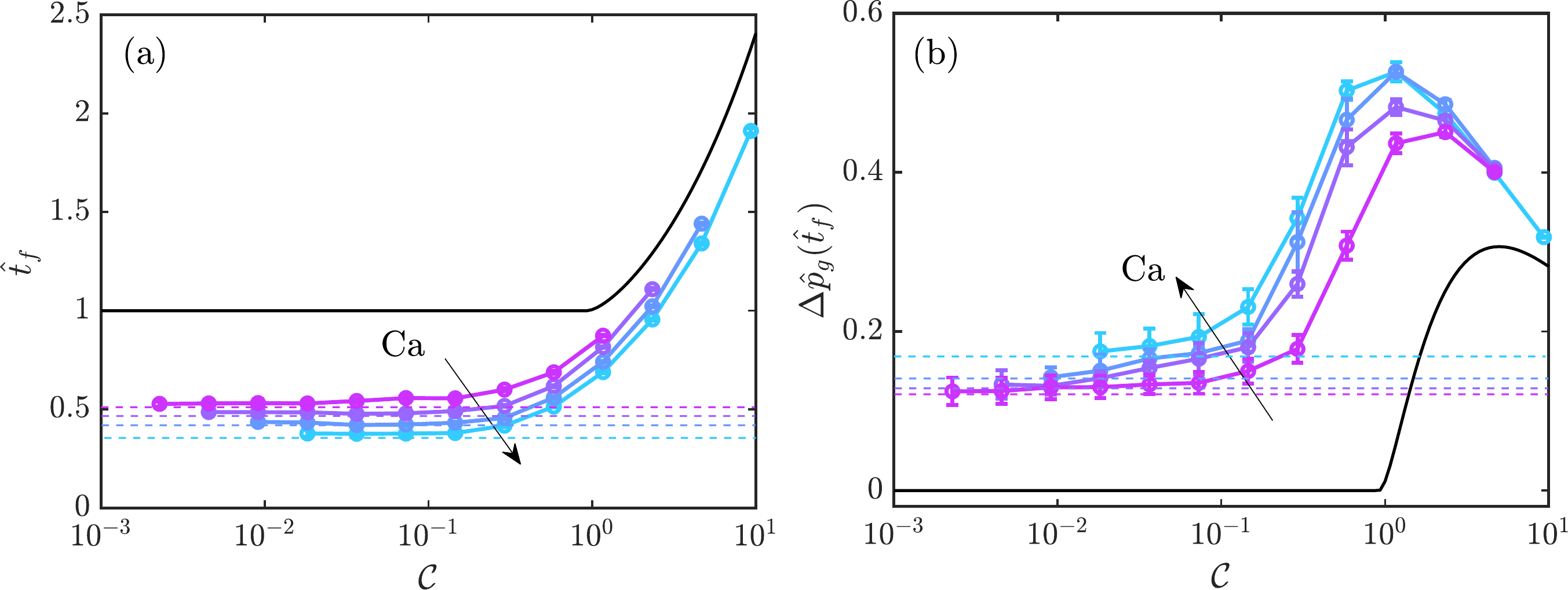}
	\caption{(a) Breakout time $\hat{t}_f$ at which the interface first reaches the rim of the flow cell and (b) corresponding breakout gauge gas pressure $\Delta\hat{p}_g(\hat{t}_f)$, both computed from numerical solution of Eq.~\eqref{eq:CompressibleODE} for $\mathcal{R} = 0.025$ (black lines) and from the full numerical simulations (coloured curves and symbols) for $\Ca\in\{2.61\times10^3$, $5.21\times10^3$, $1.04\times10^4$, $2.08\times10^4$\} (increasing in the direction of the arrows). Each point is the average value from 10 simulations, with error bars equal to one standard deviation above and below the mean. Error bars are smaller than symbols in panel~(a). The incompressible results are shown for reference (horizontal dashed lines).}
	\label{fig:BOtime}
\end{figure}

In Figures~\ref{fig:BOtime}(a, b), we plot the breakout time and the breakout pressure, respectively. We compute these values numerically from the axisymmetric model (solid black curves) for $\mathcal{R}=0.025$. We find that, for $\Comp\lesssim1$, the breakout time $\hat{t}_f$ is almost exactly 1 (to within numerical resolution) while the breakout pressure $\Delta\hat{p}_g(\hat{t}_f)$ is almost exactly zero (to within numerical resolution). This corresponds to a scenario where breakout occurs at exactly the same time as for an incompressible flow driven at the nominal injection rate. As a consequence, the volume of air displaced by the piston at the moment of breakout is exactly equal to the volume of liquid displaced by the air, such that the air returns to its initial volume and pressure. Hence, the driving compressive force vanishes at the same rate as the viscous resistive force, consistent with terminating on the trivial solution $\bar{Q}_-$. For $\Comp\gtrsim1$, in contrast, breakout is delayed ($\hat{t}_f>1$) which means that the air is still compressed at breakout and $\Delta\hat{p}_g(\hat{t}_f)>0$. Hence, the driving pressure remains finite as the opposing resistance vanishes and $\hat{Q}_b$ diverges.

For comparison, in Fig.~\ref{fig:BOtime} we also show the breakout time and the breakout pressure from simulations with fingering over a range of $\Comp$ at different values of $\Ca$ (colored curves and symbols). We observe qualitatively similar behavior, with $\hat{t}_f$ and $\Delta\hat{p}_g(\hat{t}_f)$ varying slowly for $\Comp\lesssim0.1$ before increasing sharply around $\Comp\approx0.1$. We also observe nonmonotonic variation in $\Delta\hat{p}_g(\hat{t}_f)$ for the highest $\Ca$ studied, and we would expect to see the same for the lowest $\Ca$ were the range of $\Comp$ extended. Breakout times are systematically and substantially lower in the simulations than in the axisymmetric model because a significant fraction of the liquid is bypassed by viscous fingers, allowing the interface to reach the edge of the cell earlier. Similarly, the breakout pressures are systematically greater in the fingering simulations, which may reflect the significant volume of liquid left in the cell at the moment of breakout; because the interface is still advancing, a significant pressure is still required to drive flow in the remaining liquid.

\subsection{Linear stability analysis}\label{sec:Stability}

\begin{figure}
	\centering
	\includegraphics[width=0.8\linewidth]{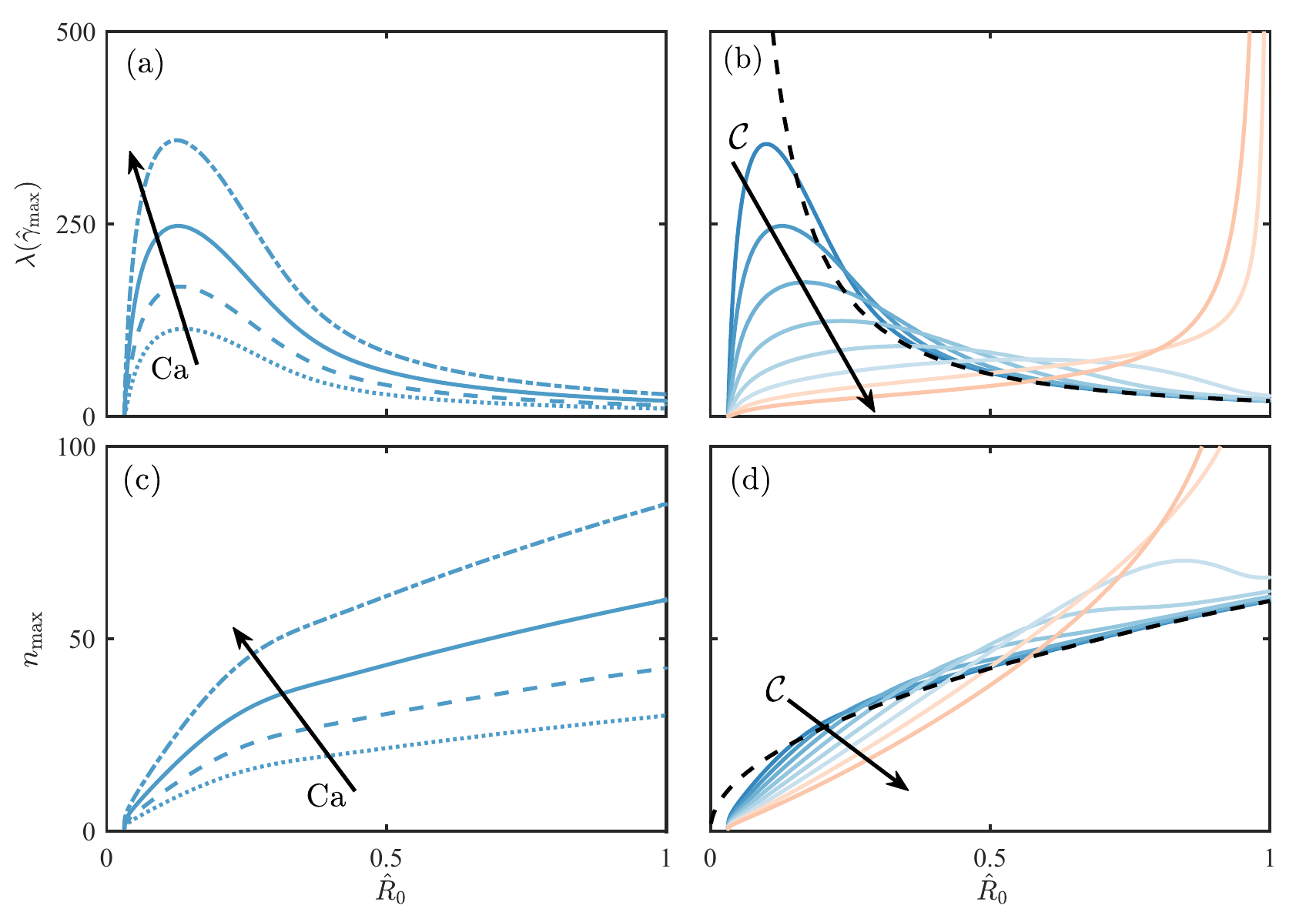}
	\caption{(a) Growth rate of the most unstable mode of perturbation [Eq.~\eqref{eq:GrowthRate}] and (c) most unstable mode of perturbation [Eq.~\eqref{eq:UnstableMode}] for $\mathcal{C} = 0.14$ and $\Ca\in\{2.61\times10^3$, $5.21\times10^3$, $1.04\times10^4$, $2.08\times10^4\}$ (increasing in the direction of the arrows). Panels (b) and (d) show the same quantities for $\Ca=1.04 \times 10^4$ and (blue to red) $\mathcal{C}\in \{0.018$, $0.036$, $0.071$, $0.14$, $0.28$, $0.57$, $1.13$, $2.27\}$ (increasing in the direction of the arrows). Blue and red shades indicate $\Comp<1$ and $\Comp>1$, respectively, and the classical incompressible values are shown for comparison (dashed black). Note that while the incompressible values are defined for all $\hat{R}_0\ge0$, compressible values are only defined for $\hat{R}_0\ge\mathcal{R}=0.025$.}
	\label{fig:LinearStabilityAnalysis}
\end{figure}

We next examine the impact of compressibility on the onset of viscous fingering by performing a linear stability analysis. To do so, we consider a slightly perturbed circular solution of the form
\begin{align}
	\hat{R}(\theta, \hat{t}) &= \hat{R}_0(\hat{t}) + \varepsilon \hat{\gamma}_n(\hat{t}) \cos n \theta + \mathcal{O}(\varepsilon^2), \label{eq:Perturbation1} \\
	\hat{p}(\hat{r}, \theta, \hat{t}) &= \hat{p}_0(\hat{r}, \hat{t}) + \varepsilon \hat{A}_n(\hat{r},\hat{t}) \cos n \theta + \mathcal{O}(\varepsilon^2), \label{eq:Perturbation2} 
\end{align}
where $\varepsilon \ll 1$, $n \ge 2$, and $\hat{R}_0$ and $\hat{p}_0$ are the unperturbed circular solution (i.e., the base state). Here, $\hat{\gamma}_n$ and $\hat{A}_n$ denote, respectively, the amplitudes of the $n$th mode of perturbation to the radius and pressure. Following \citet{Paterson1981}, we use the $\mathcal{O}(\varepsilon)$ problem to derive an evolution equation for the relative growth rate of the $n$th mode of perturbation,
\begin{align}
	\RRC{\hat{\gamma}_n} = \frac{n-1}{2\hat{R}_0^2} \left[ \hat{Q}_b - \frac{n(n+1)}{\Ca \hat{R}_0} \right] \label{eq:GrowthRate}.
\end{align}
The most unstable mode of perturbation is then
\begin{align}
	n_{\max} = \sqrt{\frac{1 + \Ca \hat{R}_0 \hat{Q}_b}{3}}, \label{eq:UnstableMode}
\end{align}
which comes about by solving $\p \RRC{\hat{\gamma}_n}/\p n = 0$. Thus, the value of $n_{\max}$ depends on the evolution of $\hat{Q}_b(\hat{R}_0)$ for an unperturbed circular interface. This base state is precisely the solution to the axisymmetric model, where $\hat{Q}_b=2\hat{R}_0\dot{R}_0$. By combining the axisymmetric model and linear stability analysis, we may therefore understand how compression-driven displacement modifies the onset of viscous fingering.

Figures~\ref{fig:LinearStabilityAnalysis}(a, c) show $\RRC{\hat{\gamma}_{\max}}$ and $n_{\max}$, respectively, as functions of $\hat{R}_0$ for fixed $\Comp$ at different values of $\Ca$. These curves are calculated by solving Eq.~\eqref{eq:CompressibleODE} numerically and substituting $\hat{Q}_b(\hat{R}_0)$ into Eqs.~\eqref{eq:GrowthRate} and \eqref{eq:UnstableMode}. Increasing $\Ca$ enhances the growth rate of the instability [$\RRC{\hat{\gamma}_n}$ increases] as well as the most unstable mode of perturbation $n_{\max}$ for all $\hat{R}_0$. These observations are consistent with our experimental and numerical results (Figs.~\ref{fig:Experiments} and \ref{fig:Simulations}) where, over the values of $\mathcal{C}$ considered, increasing $\Ca$ results in more prominent branching and tip splitting behaviour.

Fixing $\Ca$ and varying $\Comp$, as in Figs.~\ref{fig:LinearStabilityAnalysis}(b, d), we find that increasing $\Comp$ suppresses the instability at earlier times and promotes it at later times. This observation is consistent with the fact that, as discussed in section~\ref{sec:BulkQP}, compression-driven displacement is characterised by a lower injection rate at earlier times and a greater injection rate at later times, compared with the nominal injection rate. As a result, for a given $\Comp$, both $\RRC{\hat{\gamma}_n}$ and $n_{\max}$ cross the incompressible solution (dashed line; $\hat{Q}_b=1$) at some value of $\hat{R}_0$.

Physically, the instability is driven by viscous forces in the defending liquid and resisted by capillary forces at the interface. The stabilising effect of capillary forces means that sufficiently high wavenumber (i.e. short wavelength) modes decay; onset occurs as the radius of the interface reaches a critical value, at which the perimeter of the interface becomes large enough to accommodate the longest-wavelength unstable mode~\citep{Paterson1981}. In the incompressible system, where the injection rate is fixed at $\hat{Q}_b = 1$, \citet{Paterson1981} showed that this critical radius $\hat{R}_{0i}$ satisfies
\begin{align}\label{eq:rcIncomp}
	\sqrt{\Ca\hat{R}_{0i}+\frac{1}{4}} - \frac{1}{2}=2\pi.
\end{align}
Hence, $\hat{R}_{0i}\sim\Ca^{-1}$, such that onset occurs at smaller radii for larger $\Ca$ or, say, larger $Q$. In the compressible system, the time-dependent injection rate is $\hat{Q}_b=2\hat{R}_0\dot{R}_0$, where $\dot{R}_0(\hat{t})$ is given by Eq.~\ref{eq:CompressibleODE}. The critical radius $\hat{R}_{0c}$ in the compressible system must then satisfy
\begin{align}\label{eq:rcComp}
	\sqrt{\frac{2\Ca\hat{R}_{0c}}{\Comp}\left(\frac{\hat{R}_{0c}^2-\mathcal{R}^2 - \hat{t}_c}{\ln(\hat{R}_{0c})}\right)+\frac{1}{4}} - \frac{1}{2}=2\pi,
\end{align}
where $\hat{t}_c$ is the time of onset (i.e., $\hat{R}_0(\hat{t}_c) = \hat{R}_{0,c}$). To close Eq.~\eqref{eq:rcComp}, we require $\hat{R}_0(\hat{t})$, which we calculate numerically from the axisymmetric model. Hence, onset in the compressible system depends on both $\Ca$ and $\Comp$, with the latter dictating the unsteady injection rate.

\begin{figure}
	\centering
	\includegraphics[width=0.50\linewidth]{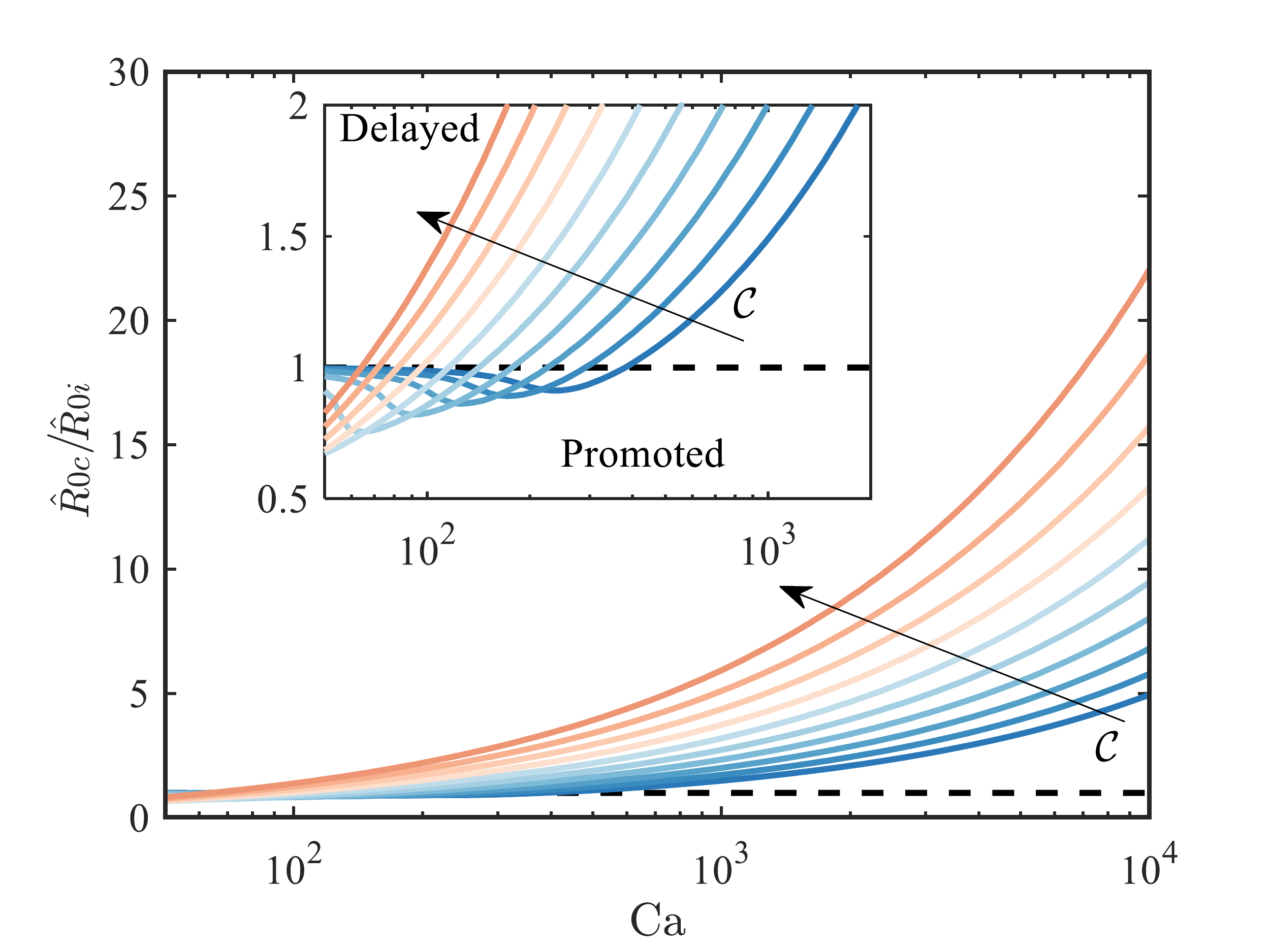}
	\caption{Onset radius $\hat{R}_{0c}$ in the compressible system [Eq.~\eqref{eq:rcComp}], normalised by the onset radius $\hat{R}_{0i}$ in the incompressible system [Eq.~\eqref{eq:rcIncomp}] at the same $\Ca$, as functions of $\Ca$ for $\Comp$ ranging from 0.01 to 10 (increasing in the direction of the arrows). Red and blue curves, respectively, indicate $\Comp>1$ and $\Comp<1$. The dashed black line denotes $\hat{R}_{0c}=\hat{R}_{0i}$. Inset: Close-up of the low-$\Ca$ region where compressibility can weakly promote onset ($\hat{R}_{0i}<\hat{R}_{0c}$), according the predictions of linear stability analysis.}
	\label{fig:LinStabOnset}
\end{figure}

In Figure~\ref{fig:LinStabOnset}, we compare the linear stability predictions of onset in the compressible and incompressible systems by plotting $\hat{R}_{0c}/\hat{R}_{0i}$ as a function of $\Ca$ for varying $\Comp$. We determine $\hat{R}_{0c}$ numerically, while $\hat{R}_{0i}$ is given analytically by Eq.~\eqref{eq:rcIncomp}. We observe a substantial delay in the onset of the instability, indicated by $\hat{R}_{0c}/\hat{R}_{0i}>1$, when $\Ca\gtrsim10^2$. At the highest $\Ca$ and $\Comp$ shown, onset is delayed in the compressible system to radii more than 20 times greater than in the incompressible system at the same $\Ca$. The strong delaying effect derives from the very low initial injection rates $\hat{Q}_b\ll1$ [see Fig~\ref{fig:VolPressureComp}(a)] associated with compression-driven displacement. The effect is amplified with increasing $\Comp$, which leads to lower initial $\hat{Q}_b$, and with increasing $\Ca$, which corresponds to smaller onset radii in the incompressible system. Hence, if onset is predicted at small radii $\hat{R}_0\ll1$ for an incompressible flow, then the slow initial injection rate can significantly delay onset in a compression-driven flow. The inset of Fig.~\ref{fig:LinStabOnset} shows a magnified plot of $\hat{R}_{0c}/\hat{R}_{0i}$ at $\Ca\lesssim10^3$. For low $\Ca$, linear stability predicts that compressibility may act to promote the onset of viscous fingering relative to an incompressible system, such that $\hat{R}_{0c}/\hat{R}_{0i}<1$. This promoting effect at low $\Ca$ is due to the relatively high injection rates ($\hat{Q}_b>1$) at later times or larger $\hat{R}_0$, which linear stability analysis suggests may trigger onset at smaller radii relative to an incompressible flow. This promoting effect is much weaker than the delaying effect of high $\Ca$, reducing the radius of onset by less than a factor of 2.  Moreover, the effect is triggered for large $\hat{R}_0\lessapprox1$, and it is therefore unclear whether the effect on viscous fingering would be visible before breakout.

\subsection{Nonlinear finger growth} \label{sec:Nonlinear}

We now return to fully nonlinear pattern formation in both experiments [Fig.~\ref{fig:Experiments}] and simulations [Fig.~\ref{fig:Simulations}]. We first demonstrate that the systematic delay in the onset of fingering predicted by linear stability analysis due to compressibility at high $\Ca$ is readily observable in both experiments and simulations. We then examine the impact of this delayed onset on the resulting fingering pattern. We also examine the agreement between simulations and experiments, both in terms of the point of onset and the details of the resulting pattern. To quantify the severity of the fingering pattern, we consider the isoperimetric ratio
\begin{align}
	\mathcal{I} = \frac{L^2}{4 \pi A}, \label{eq:Isoperimetric}
\end{align}
where $L$ and $A$ are, respectively, the length of the interface and the area it encloses. For a circular interface, $\mathcal{I} = 1$. Hence, any deviation from unity indicates some perturbation away from the axisymmetric base state assumed in the linear stability analysis [\S~\ref{sec:Stability}].

\begin{figure}
	\centering
	\includegraphics[width=0.775\linewidth]{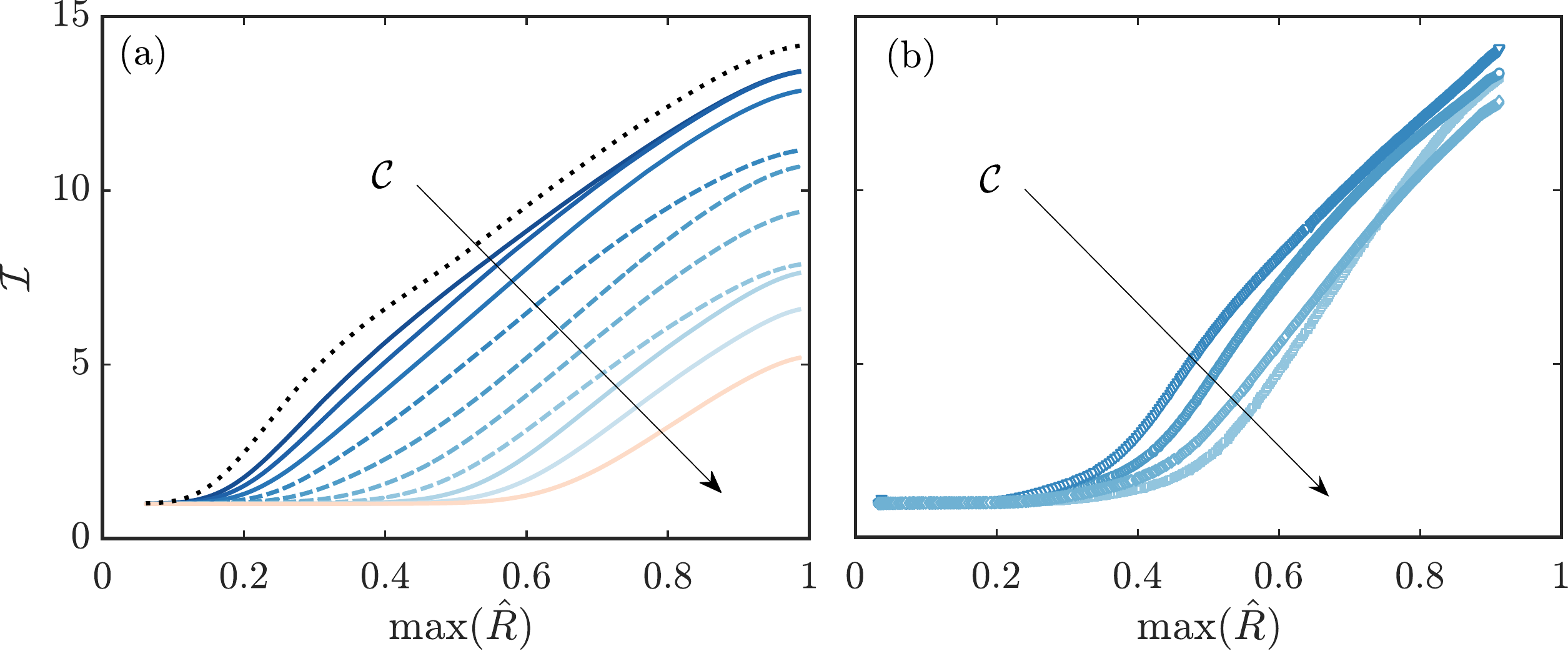}	
	\caption{Isoperimetric ratio $\mathcal{I}$ [Eq.~\eqref{eq:Isoperimetric}] as a function of the maximum radial extent of the interface at $\Ca = 5.21 \times 10^{3}$ from (a) numerical simulations at $\mathcal{C} \in\{4.5\times10^{-3},$ 0.009, 0.018, 0.036, 0.071, 0.14, 0.28, 0.57, 1.13, 2.26\}, and (b) experiments at $\mathcal{C}\in \{$0.036, 0.071, 0.14, 0.28\} ($Q=2.5$~mL/min, $V_g(0)\in\{25,50,100,200\}$~mL). The incompressible solution is shown in (a) for comparison (dotted black). Dashed curves in (a) are at the values of $\Comp$ used for the experiments in (b).}
	\label{fig:metrics}
\end{figure}

Figure~\ref{fig:metrics} shows the evolution of $\mathcal{I}$ at fixed $\Ca$ for a range of $\Comp$ for simulations [Figs.~\ref{fig:metrics}(a)] and experiments [Figs.~\ref{fig:metrics}(b)] as functions of the maximal radial extent of the interface $\max(\hat{R})$. These experiments and simulations correspond to the top row of Figures~\ref{fig:Experiments} and \ref{fig:Simulations}, respectively. The isoperimetric ratio $\mathcal{I}$ is initially close to 1, suggesting that the interface is near-circular, before increasing monotonically and at a relatively steady rate for the remainder of the experiment or simulation, corresponding to the growth of viscous fingers. The departure from $\mathcal{I}\approx1$ occurs at larger radii [$\max(\hat{R})$] for larger $\Comp$ (arrows) in both simulations and experiments, consistent with a delayed onset. The growth of $\mathcal{I}$ with $\max(\hat{R})$ is somewhat faster in experiments than in simulations, consistent with the more frequent occurrence of tip-splitting and side-branching visible in the experimental images~(Fig.~\ref{fig:Experiments}).

\begin{figure}
	\centering
	\includegraphics[width=0.8\linewidth]{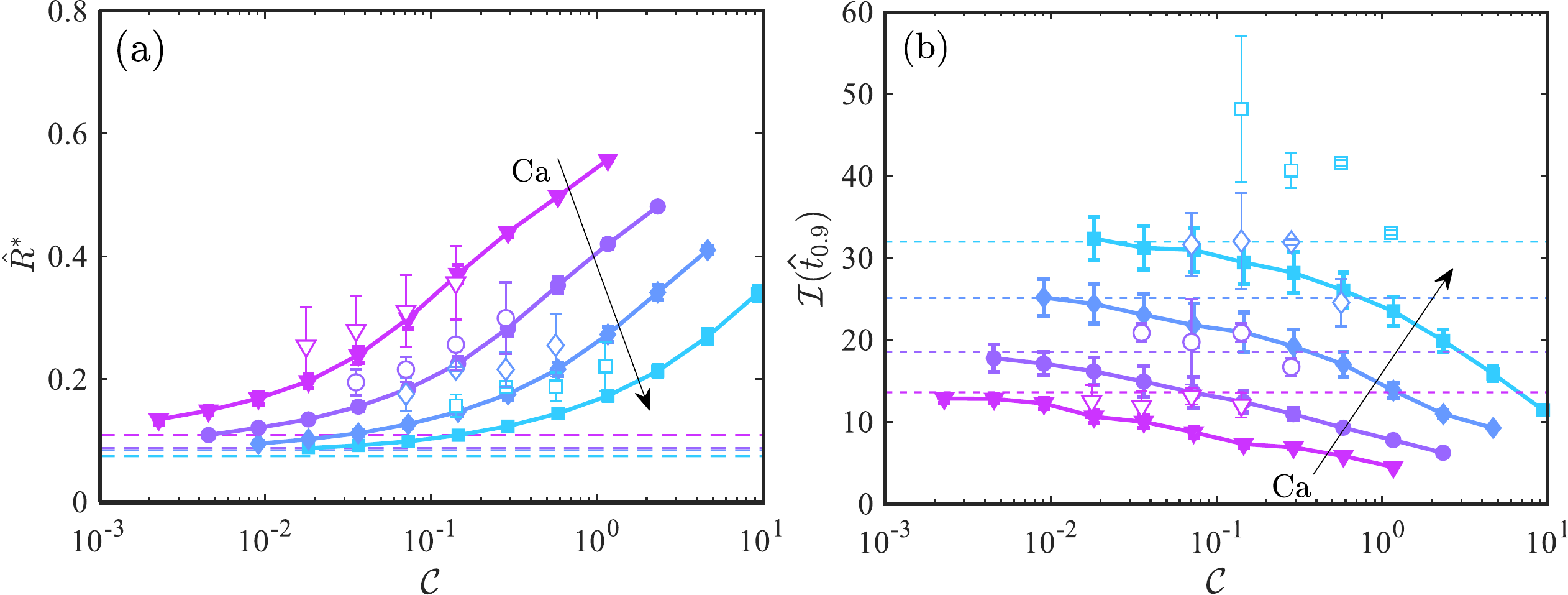}
	\caption{\label{fig:metricsFin}(a) Onset radius $\hat{R}^*$ at which $\mathcal{I}$ first exceeds 1.1 (c.f. Fig.~\ref{fig:metrics}) and (b) the near-breakout isoperimetric ratio $\mathcal{I}(\hat{t}_{0.9})$ measured at $\hat{t}_{0.9}=\hat{t}(\max(\hat{R})=0.9)$ as functions of $\Comp$ for a range of $\Ca$. Symbols and colours correspond to $\Ca=2.61\times10^3$, $5.21\times10^3$, $1.04\times10^4$, and $2.08\times10^4$ with arrows indicating increasing $\Ca$. Simulation results and experimental results are plotted with small connected symbols and large scattered symbols, respectively. The incompressible results are shown for reference (horizontal dashed lines).}
\end{figure}

To quantitatively examine how $\Comp$ impacts the onset of fingering, we define the radius at onset $\hat{R}^*$ in our experiments and simulations as being $\max(\hat{R})$ at the last recorded instant (video frame or time step) for which $\mathcal{I}<1.1$. The measured and computed values of $\hat{R}^*$ as functions of $\Comp$ for different $\Ca$ are shown in Fig.~\ref{fig:metricsFin}(a) for experiments (scattered symbols) and simulations (connected symbols). Qualitatively, both sets of data show the same behaviour: the onset radius $\hat{R}^*$ tends to increase with increasing $\Comp$ or decreasing $\Ca$. The latter result is familiar from classical studies of viscous fingering \cite{Paterson1981}. Both results are consistent with the linear stability analysis presented in \S~\ref{sec:Stability}, confirming the prediction that increasing $\Comp$ delays the onset of fingering. Moreover, varying $\Ca$ or $\Comp$ by a similar amount (an order of magnitude, say) yields a comparable (and opposite) effect on the radius of onset. Comparing experiments and simulations directly, there is reasonable quantitative agreement between the two over the experimental parameter range studied: the majority of experimental data points lie within one standard deviation of the corresponding simulation. The gradient of the experimental data, however, appears shallower than that of the simulations, which suggests that increasing $\Comp$ is less effective in delaying onset in experiments.

The broad range of patterns generated by the fingering instability after onset can be quantified by considering $\mathcal{I}(\hat{t}_{0.9})$. That is, the isoperimetric ratio at the near-breakout time $\hat{t}_{0.9}=\hat{t}(\max(\hat{R})=0.9)$ (see \S~\ref{sec:DataProc}). We plot $\mathcal{I}(\hat{t}_{0.9})$ as a function of $\Comp$ for different values of $\Ca$ in Fig.~\ref{fig:metricsFin}(b). In simulations, we find that $\mathcal{I}(\hat{t}_{0.9})$ tends to the incompressible case (dashed lines) as $\mathcal{C} \to 0$, and the severity of the fingering pattern at $\hat{t}_{0.9}$ decreases with increasing $\Comp$ or decreasing $\Ca$. Qualitatively, this behaviour is consistent with experiments, though experimental measurements are systematically greater, again consistent with the observed prevalence of tip-splitting and side-branching in experiments~(Fig.~\ref{fig:Experiments}).

As discussed in \S~\ref{sec:Exp} and \S~\ref{sec:BulkQP}, the systematically greater pressures observed in experiments than in simulations may be in part due to the qualitatively and quantitatively different fingering patterns. For instance, though $\mathcal{I}$ increases more steeply in experiments than in simulations (Fig.~\ref{fig:metrics}), indicating more severe fingering which was previously attributed to lower pressures, the delayed onset in experiments [Fig.~\ref{fig:metricsFin}(a)] could counter this effect due to the need to displace more liquid at earlier times. Furthermore, the greater instances of side-branching and tip-splitting in experiments may also correspond to greater pressures due to the displacement of liquid between the primary fingers. The exact correspondence between finger morphology and driving pressure is beyond the scope of this study, but merits future investigation.

\begin{figure}
	\centering
	\includegraphics[width=0.8\linewidth]{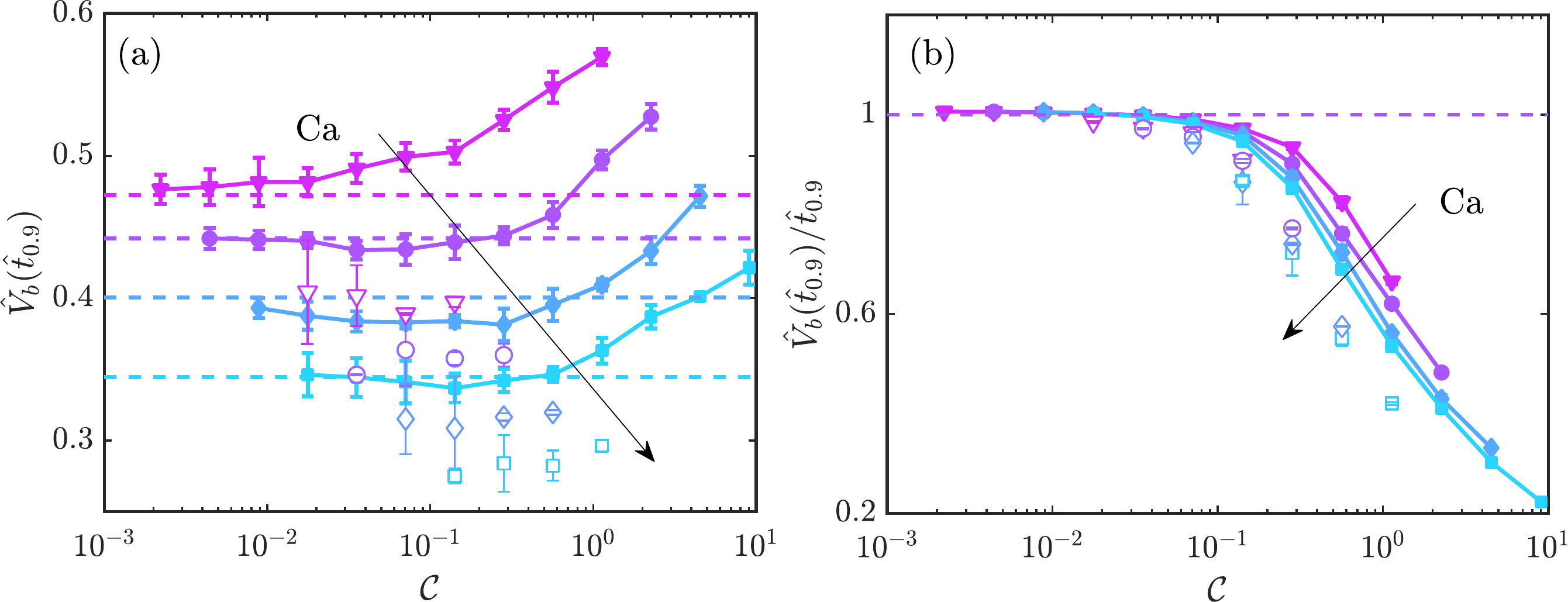}
	\caption{\label{fig:metricsQbar} (a) Volume of air in the flow cell at near-breakout, $\hat{V}_b(\hat{t}_{0.9})$, and (b) overall average injection rate up to near-breakout, $\hat{V}_b(\hat{t}_{0.9})/\hat{t}_{0.9}$, as functions of $\Comp$ for a range of $\Ca$. Note that $\hat{V}_b(\hat{t}_{0.9})$ is also approximately equal to the near-breakout volume of liquid expelled since $\hat{V}_b(0)\ll{}\hat{V}_b(\hat{t}_{0.9})$. Symbols and colours correspond to $\Ca=2.61\times10^3$, $5.21\times10^3$, $1.04\times10^4$, and $2.08\times10^4$ with arrows indicating increasing $\Ca$. Simulation results and experimental results are plotted with small connected symbols and large scattered symbols, respectively. The incompressible results are shown for reference (horizontal dashed lines).}
\end{figure}

Finally, we return to the bulk displacement dynamics to shed some light on the increasingly delayed onset and mitigated fingering with $\Comp$. The near-breakout volume of air in the cell, $\hat{V}_b(\hat{t}_{0.9})$, which is approximately equal to the near-breakout volume of liquid expelled since $\hat{V}_b(0)\ll{}\hat{V}_b(\hat{t}_{0.9})$, is close to the incompressible value and essentially independent of $\Comp$ for $\Comp\lesssim0.1$ [Fig.~\ref{fig:metricsQbar}(a)]. Recall that the same is true of the breakout time $\hat{t}_f$ [Fig.~\ref{fig:BOtime}(a)]. Thus, the \textit{average} injection rate up to near-breakout (\textit{i.e.}, the average value of $\hat{Q}_b$ during the interval $0\leq\hat{t}\leq\hat{t}_{0.9}$), which is given by $\hat{V}_b(\hat{t}_{0.9})/\hat{t}_{0.9}$, is approximately unity for $\Comp\lesssim0.1$ [Fig.~\ref{fig:metricsQbar}(b)]. That is, the average injection rate is approximately equal to the nominal injection rate for $\Comp\lesssim0.1$, despite the non-trivial time evolution of the instantaneous injection rate $\hat{Q}_b(t)$. Nonetheless, these smaller values of $\mathcal{C}$ lead to noticeably delayed onset and reduced intensity of fingering [Fig.~\ref{fig:metricsFin}]. Both $\hat{V}_b(\hat{t}_{0.9})$ and $\mathcal{I}(\hat{t}_{0.9})$ increase substantially as $\Comp$ increases further ($\Comp\gtrsim{}0.1$) [Figs.~\ref{fig:metricsFin}b and \ref{fig:metricsQbar}b], but $\hat{V}_b(\hat{t}_{0.9})/\hat{t}_{0.9}$ decreases substantially [Fig.~\ref{fig:metricsQbar}]. Thus, weak compressibility ($\mathcal{C}\lesssim{}0.1$) delays onset and mitigates fingering by introducing a time-varying $\hat{Q}_b(\hat{t})$ while roughly preserving the average injection rate, whereas stronger compressibility ($\mathcal{C}\gtrsim{}0.1$) further delays onset, mitigates fingering, and increases the volume of liquid expelled by reducing the average injection rate.

Note that we observe systematically lower values of $\hat{V}_b(\hat{t}_{0.9})$ in experiments than in simulations [Fig.~\ref{fig:metricsQbar}a], which is consistent with the systematically larger values of $\mathcal{I}(\hat{t}_{0.9})$ [Figs.~\ref{fig:metricsFin}b] and further suggests that the experiments are subject to more severe fingering than the simulations at the same values of $\mathrm{Ca}$ and $\mathcal{C}$.

\section{Discussion and conclusions} \label{sec:Discussion}

We have studied gas--liquid displacement in a rigid Hele-Shaw cell, driven by the steady compression of a connected gas reservoir. By considering an axisymmetric interface, we developed a simple axisymmetric model [Eq.~\eqref{eq:CompressibleODE}] analogous to the recent work of \citet{Cuttle2023a}, who studied compression-driven displacement in a capillary tube. The unsteady injection rate and the gas pressure in the axisymmetric model are controlled by a single dimensionless parameter, the compressibility number $\Comp$, and are independent of the capillary number $\Ca$. Remarkably, in experiments and simulations, which are subject to viscous fingering and therefore strongly non-axisymmetric, we found that the injection rate and gas pressure were still controlled primarily by $\Comp$. Variations in $\Ca$ had a far more modest effect on these `bulk' dynamics, despite having a strong influence on the severity of the fingering instability. We therefore argue that $\Comp$ is the key control parameter for bulk displacement dynamics, even for hydrodynamically unstable flows.

The axisymmetric model [Eq.~\eqref{eq:CompressibleODE}] also revealed two underlying dynamical regimes that arise from the basic coupling between a viscous displacement flow and the volumetric compression of a gas. The low- and high-$\Comp$ regimes correspond to ``on-time'' and quasisteady or delayed and burst-like expulsion at the moment of breakout, when the interface reaches the outlet of the cell. We rationalised these regimes by following the dynamical-systems approach employed by \citet{Cuttle2023a} in studying the corresponding capillary-tube problem. As in the capillary tube, there exists a critical compressibility number $\Comp=1$, which dictates the transition between quasi-steady and burst-like dynamics. In our axisymmetric model, $\Comp$ plays a directly analogous role, with the trivial solutions of the system vanishing for $\hat{R}_0\le1$ at $\Comp=1$. However, due to the evolving base state (increasing radius) of the axisymmetric model, we identified an additional sensitivity to the initial radius $\mathcal{R}$ that can influence whether the dynamics are quasi-steady or burst-like. While the axisymmetric model neglects the fingering instability, we nonetheless found that the delayed breakout and over-pressure predicted by the model were qualitatively recovered in simulations (and experiments; not shown) [Fig.~\ref{fig:BOtime}], which again points to the robust role of compression-driven displacement dynamics in a pattern-forming system.

To understand the impact of compression-driven displacement dynamics on the onset of viscous fingering, in \S~\ref{sec:Stability}, we performed a linear stability analysis of the axisymmetric model, which we took as the base state. We found that the growth rate and the most unstable mode of the perturbations depended strongly on both $\Ca$ and $\Comp$. Specifically, $\Ca$ modulates the relative strengths of the destabilising viscous and stabilising capillary forces at a given flow rate, as in the classical system, while $\Comp$ sets the evolution of the time-dependent injection rate. Our analysis predicted that, compared to an incompressible flow at the same nominal injection rate $Q$, compressibility may either suppress or promote the onset of fingering, depending on $\Ca$. The promoting effect is relatively weak and would occur at very low $\Ca$, an order of magnitude lower than the experiments presented in this work, so it is uncertain whether one could observe its effects in practice. In contrast, the delaying effect at high $\Ca$ is much more pronounced and is indeed readily observed in both experiments and simulations. In fact, increasing $\Comp$ was found to be as effective in delaying onset as decreasing $\Ca$ by a similar factor. 
The result of this delay is that the severity of the fingering pattern, as measured by the isoperimetric ratio [Eq.~\eqref{eq:Isoperimetric}], decreases substantially as $\Comp$ increases at fixed $\Ca$. This mitigated finger growth can be largely attributed to the delayed onset predicted by linear stability analysis.

In the context of viscous fingering with incompressible fluids, numerous studies have considered the impact of imposing a time-varying injection rate $Q_b(t)$, typically with the goal of identifying the $Q_b(t)$ that minimises or otherwise controls the number of fingers that develop. For example, one pair of studies considered strategies to suppress fingering by varying $Q_b(t)$ while keeping the average injection rate constant (\textit{i.e.}, while still injecting a given total volume in the same time total time)~\citep{Dias2010, Dias2012}. \citet{Dias2010} found that a piecewise-constant $Q_b(t)$ with a small initial rate followed by a larger subsequent rate was effective in suppressing onset, whereas \citet{Dias2012} found that the optimal form of $Q_b(t)$ was linearly increasing in time. Although compressibility leads to a natural and passive (rather than actively controlled) variation in the injection rate, our results share several features with these previous works. Specifically, the time-varying rates $Q_b(t)$ observed here for $\Comp\lesssim0.1$ [Fig.~\ref{fig:VolPressureComp}a,b] mimic a small-to-large variation that preserves the average rate, as suggested by~\citet{Dias2010} [Fig.~\ref{fig:metricsQbar}(b)]. In addition, the rates observed here for $\Comp\gtrsim0.1$ mimic the monotonically increasing rates suggested by~\citet{Dias2012}. Thus, although we have not specifically investigated optimisation here, our results suggest that compression-driven displacement at $\Comp\approx0.1$ would passively achieve the strongest delay in onset while preserving the nominal injection rate on average, and would therefore be optimal (in the sense of \citep{Dias2010}) at a given $\Ca$.

By conducting extensive experimental and numerical studies in tandem, we were able to thoroughly compare state-of-the-art simulations with physical data. The most impressive agreement between the two was found in the volume growth rate of the bubbles and, to a lesser extent, the evolution of the pressure. This difference is despite the broad variation in patterns, quantified by the isoperimetric ratio, observed in experiments and simulations at the same parameters. Indeed, even our repeat experiments were subject to significant variability in fingering behaviour, and yet were highly reproducible in terms of injection rate and pressure (Appendix~\ref{app:reprod}). These observations speak to the robust nature of the underlying dynamics of compression-driven displacement that modulate the growth of the interface, and which can be described to leading order by the single parameter $\Comp$.

There are several possible sources for the differences in pattern formation (\textit{e.g.}, the isoperimetric ratio or the qualitative interface evolution) between experiments and simulations. One such source is the choice of initial condition in the simulations [Eq.~\eqref{eq:InitialCondition}]. We chose an initial radius $r_0$ to best match with experiments. However, small variations in $r_0$, along with the choice of $\varepsilon$ (initial amplitude of perturbations) and the modes of perturbations used could have a non-negligible influence on the final shape of the interface. Further, the experiments are subject small disturbances due to plate defects, microscopic contaminants, thermal fluctuations, and the outlet conditions, amongst other culprits, which are not captured by our model. Despite being small, such disturbances can significantly modify the fingering pattern, particularly at higher $\Ca$ where the interface is far more distorted. For example, it has been shown that finite perturbations in Hele-Shaw cells and channels can exert a strong influence on pattern formation~\citep{Couder1986,Zocchi1987,Rabaud1988,Thompson2014,Zhang2021}. As our simulations and experiments are subject to very different sources of perturbations, it is unsurprising that they should produce quantitatively and qualitatively different patterns. These details, however, do not detract from the key result confirmed by both approaches; increasing the compressibility number is as effective in delaying onset as decreasing the capillary number. Compressibility acts passively in two-phase gas-driven flows and is as natural to the system as viscosity or surface tension. This is in contrast with previous control strategies~\citep{Nagatsu2007,AlHousseiny2012,PihlerPuzovic2012,Zheng2015,Gao2019} that, although effective, are often awkward to implement in practice due to the restrictions placed on the confining geometry or the choice of fluids. The compressibility number, meanwhile, is a parameter that can easily be tuned, without having to alter the system geometry, compliance, or fluid properties; simply selecting a larger syringe is sufficient. Our results therefore strongly point to compressibility as a second key parameter in the assessment and control of viscous fingering in real systems, as discussed in more detail in a companion study~\citep{Morrow2023b}. As a final remark, gas compression and the associated unsteady flows will continue to be a source of frustration in many practical and experimental settings, where steady flows are required. For those wishing to avoid such effects, our study offers a complete framework that accounts for all relevant parameters.

\textit{Data availability} --- The supporting data for this study are openly available on Zenodo \citep{cuttle-zenodo-2023}.

\textit{Published version} --- This article is published as C. Cuttle, L. C. Morrow, and C. W. MacMinn. Compression-driven viscous fingering in a radial Hele-Shaw cell. \emph{Physical Review Fluids}, 8:113904, 2023.

\begin{acknowledgments}
	We are grateful to Mr. Clive Baker for technical support. This work was supported by the European Research Council (ERC) under the European Union’s Horizon 2020 Programme [Grant No. 805469], by the UK Engineering and Physical Sciences Research Council (EPSRC) [Grant No. EP/S034587/1], and by the John Fell Oxford University Press Research Fund [Grant No. 132/012].
\end{acknowledgments}

\appendix
\section{Isothermal and adiabatic models of gas compression}\label{app:adiabatic}

In our experiments, we assumed that gas compression was isothermal, corresponding to `rapid' equilibration with the environment due to heat diffusion through the walls of the syringe. In practice, the syringe walls are borosilicate glass, which is a poor thermal conductor, and our assumptions therefore require some justification. For instance, one may calculate the typical timescale of thermal diffusion $\tau_D=W^2/\alpha_g\approx7$~s, based on the thickness of the syringe walls $W=2.2$~mm and the thermal diffusivity of borosilicate glass $\alpha_g\approx0.6$~mm$^2$/s. Alternatively, if we consider thermal diffusion in the air, which has a diffusivity of $18$~mm$^2$/s and a lengthscale on the order of the syringe barrel inner radius $16.3$~mm, we compute a similar timescale of around 14~s. We can then compute the Fourier number $\mathrm{Fo}=t_f/\tau_D$, which compares the total time $t_f$ of an experiment to the timescale of thermal diffusion. Hence, when $\mathrm{Fo}\gg1$, we expect any heat generated to dissipate rapidly to the environment and maintain a near-constant temperature during compression (i.e., an isothermal process). We find that, for $Q=[1.25,2.5,5,10]$~mL/min, $\mathrm{Fo}\approx[40,20,10,5]$, with some scatter around these mean values due to experimental variability and compressibility effects. Hence, we expect compression to be approximately isothermal for all but the fastest experiments (largest $Q$), where $\mathrm{Fo}\gtrsim1$ and compression may result in non-negligible changes in temperature.

\begin{figure}
	\centering
	\includegraphics[width=0.8\linewidth]{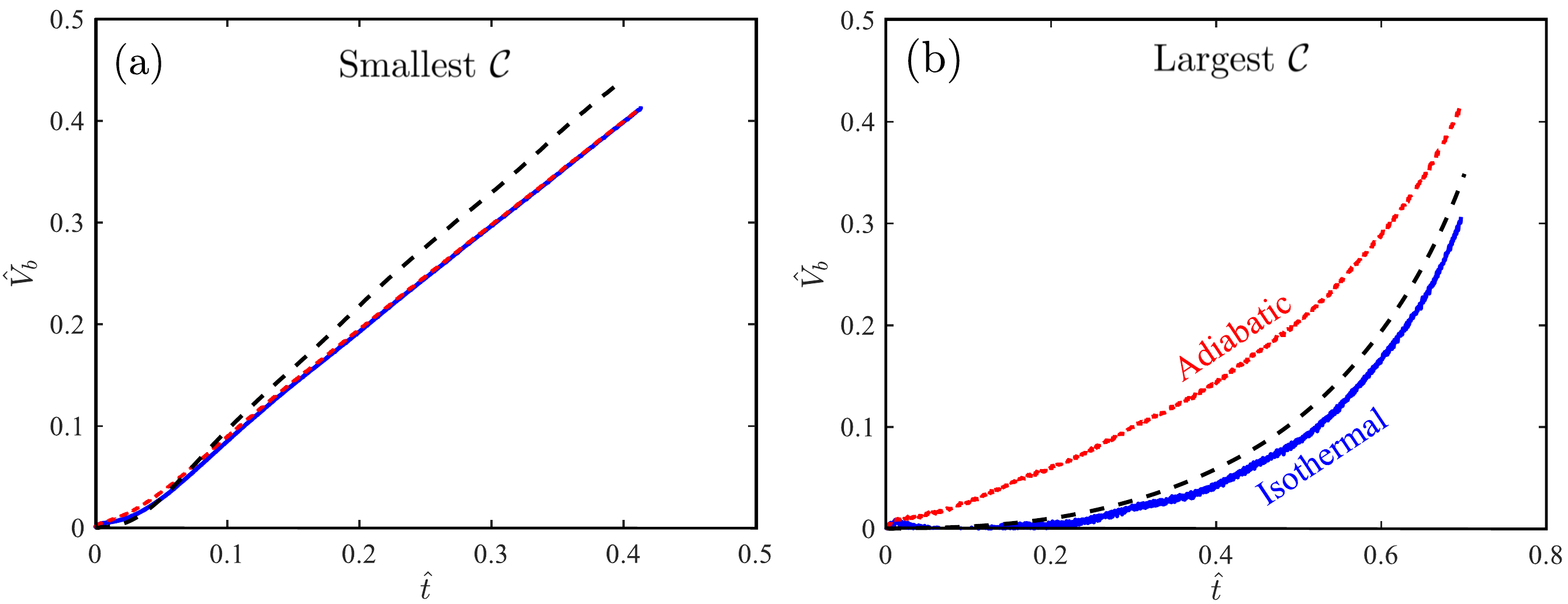}
	\caption{The normalised total experimental gas volume $\hat{V}_g$ as a function of normalised time $\hat{t}$ for isothermal (solid blue; \ref{eq:VolBub}) and adiabatic (dotted red; \ref{eq:VolBubAdiab}) models of gas compression. Dashed black curves are the normalised projected volume $\hat{A}(\hat{t})$ of the fingering pattern, which is an upper bound on $\hat{V}_g$. Panels (a) and (b) show data from experiments performed at the smallest and largest compressibility numbers studied, $\Comp=0.018$ ($Q=1.25$~mL/min, $V_g(0)=25$~mL) and $\Comp=1.13$ ($Q=10$~mL/min, $V_g(0)=200$~mL), respectively.}
	\label{fig:adiabatic}
\end{figure}

Rather than assume the gas compression is isothermal, we may have instead assumed that the compression is adiabatic, such that the system is perfectly insulated and no heat is conducted to the environment. (In reality, the process will lie somewhere between these two extremes, but a mixed model is unnecessarily complicated and would cloud the key results of our study.) To compare the two assumptions, we can compute the gas volume from the recorded gas gauge pressure (as discussed in \S~\ref{sec:DataProc}) for an adiabatic process, for which $p_g V_g^\eta$ is a constant. The adiabatic index $\eta=7/5$ for diatomic gases, of which air is predominantly composed. Equation~\eqref{eq:VolBub} then becomes
\begin{align}\label{eq:VolBubAdiab}
V_b(t)=V_b(0) + Qt-V_{\res}(0)\left[1-\left(\frac{p_\atm}{p_\atm +\Delta p_g(t)}\right)^{1/\eta}\right].
\end{align}
Figure~\ref{fig:adiabatic} compares isothermal~\eqref{eq:VolBub} and adiabatic~\eqref{eq:VolBubAdiab} calculations of the normalised bubble volume $\hat{V}_b$ as functions of normalised time $\hat{t}$ for the smallest nominal injection rate $Q$ and reservoir volume $V_\res(0)$ [Fig.~\ref{eq:VolBubAdiab}(a)] and the largest $Q$ and $V_\res(0)$ [Fig.~\ref{eq:VolBubAdiab}(b)], corresponding to the smallest and largest compressibility numbers, respectively. On the one hand, when $\Comp\propto QV_\res(0)$ is smaller, the difference between adiabatic and isothermal models is minimal [Fig.~\ref{eq:VolBubAdiab}(a)] because the pressure $\Delta p_g$ required to drive the flow is smaller and the volume of the bubble $V_b$ is more comparable to that of the reservoir $V_\res$. Thus, relative changes in the total volume of the gas are smaller and have less impact on the volume of gas in the cell. On the other hand, when $\Comp$ is larger, the two models predict significantly different $\hat{V}_b$ for the same $\Delta p_g$ [Fig.~\ref{fig:adiabatic}(b)]. Because $\mathrm{Fo}\gtrsim1$ for the largest $Q$, it is unclear which of the models is most suitable. However, we can compare the data to an independent upper bound on $V_b$, which we calculate from images of the expanding fingering pattern, multiplying the projected area of the pattern $A(t)$ by the depth of the cell $b$. The projected volume $A(t)b$ is then the maximum volume that the air could occupy in the cell at time $t$ in the absence of residual films. Comparing predictions of $\hat{V}_b$ from the isothermal and adiabatic models against the normalised projected volume $\hat{A}=Ab/(\pi R_c^2 b)$ (black dashed line), we see in Fig.~\ref{fig:adiabatic}(b) that only the isothermal model stays consistently below this upper bound, while the adiabatic model prediction far exceeds it. Hence, the isothermal model is the better choice for our experiments, even for modest $\mathrm{Fo}$.

\section{Injection from a compressed gas source}\label{app:cylinder}
In the main text, we considered injection from a syringe pump, where a fixed mass of gas is compressed at a constant volumetric rate $Q$. Here, we consider a second scenario, also commonly used in laboratories: injection from a compressed gas source via a needle valve. For brevity, we consider a circular front and neglect thin films, as in the derivation of the axisymmetric model presented in \S~\ref{sec:RedModel}.

This injection scenario corresponds to a constant mass flow of air into a reservoir of fixed volume, such that the number of moles $n(t)$ of gas in the system increases at a fixed rate $\mathrm{d}n/\mathrm{d}t=\dot{n}$. In practice, the volumetric rate of gas emitted from the needle valve is calibrated and set at $Q$ while venting into fixed atmospheric pressure $p_\atm$, such that
\begin{equation}\label{eq:Qn}
Q=\left.\frac{\mathrm{d} V_g}{\mathrm{d} t}\right|_{p_g=p_\atm}=\frac{R_gT}{p_\atm}\dot{n},
\end{equation}
where $R_g$ and $T$ are the ideal gas constant and gas temperature, respectively. However, compression of the gas once it is diverted into the flow cell means that while $\dot{n}$ may remain constant, the actual injection rate $Q_b$ will no longer be equal to $Q$. From the ideal gas law, we have $p_g(t)V_g(t)=[n(0)+\dot{n}t]R_gT$, where $V_g(t)$ is the total volume of gas in the reservoir and the cell. Also, $p_g(0)V_g(0)=n(0) RT$, so the pressure of the gas is
\begin{equation}\label{eq:Pbn}
p_{g}(t) = \frac{V_g(0)(p_\atm+\gamma\{\pi/[4 R_0(0)]+2/b\})+Qtp_\atm}{V_g(0)+V_b(t)-V_b(0)},
\end{equation}
where the initial gas pressure $p_g(0)$ is the same as in \S~\ref{sec:fullModel}. Applying the non-dimensionalisation of Eq.~\eqref{eq:NonDimen} yields
\begin{equation}\label{eq:GCnondim}
\Delta\hat{p}_g(\hat{t})=\frac{1}{\mathcal{V}+\hat{R}_0^2(\hat{t}) - \mathcal{R}^2}\left\{\mathcal{V}\left[\mathcal{P} + \frac{1}{\Ca}\left(\frac{\pi}{4\mathcal{R}}+2\alpha\right)\right]+\mathcal{P}\hat{t}\right\}-\mathcal{P}.
\end{equation} 
Finally, taking the limits $\mathcal{V} \gg1$, and $\mathcal{P}\gg(\pi/(4\mathcal{R})+2\alpha)/\Ca$, we have
\begin{equation}\label{eq:GCred}
\Delta\hat{p}_g(\hat{t})=\frac{2(\hat{t}-\hat{R}_0^2-\mathcal{R}^2)}{\Comp}.
\end{equation}
Comparing with Eq.~\eqref{eq:CompressibleODE}, we find that in the limit of the axisymmetric model, compressed gas injection is identical to syringe pump injection. We therefore expect similar dynamics in both scenarios in practice, as was recently observed by~\citet{Peng2022}.

\section{Experimental reproducibility}\label{app:reprod}

\begin{figure}
	\centering
	\includegraphics[width=1\linewidth]{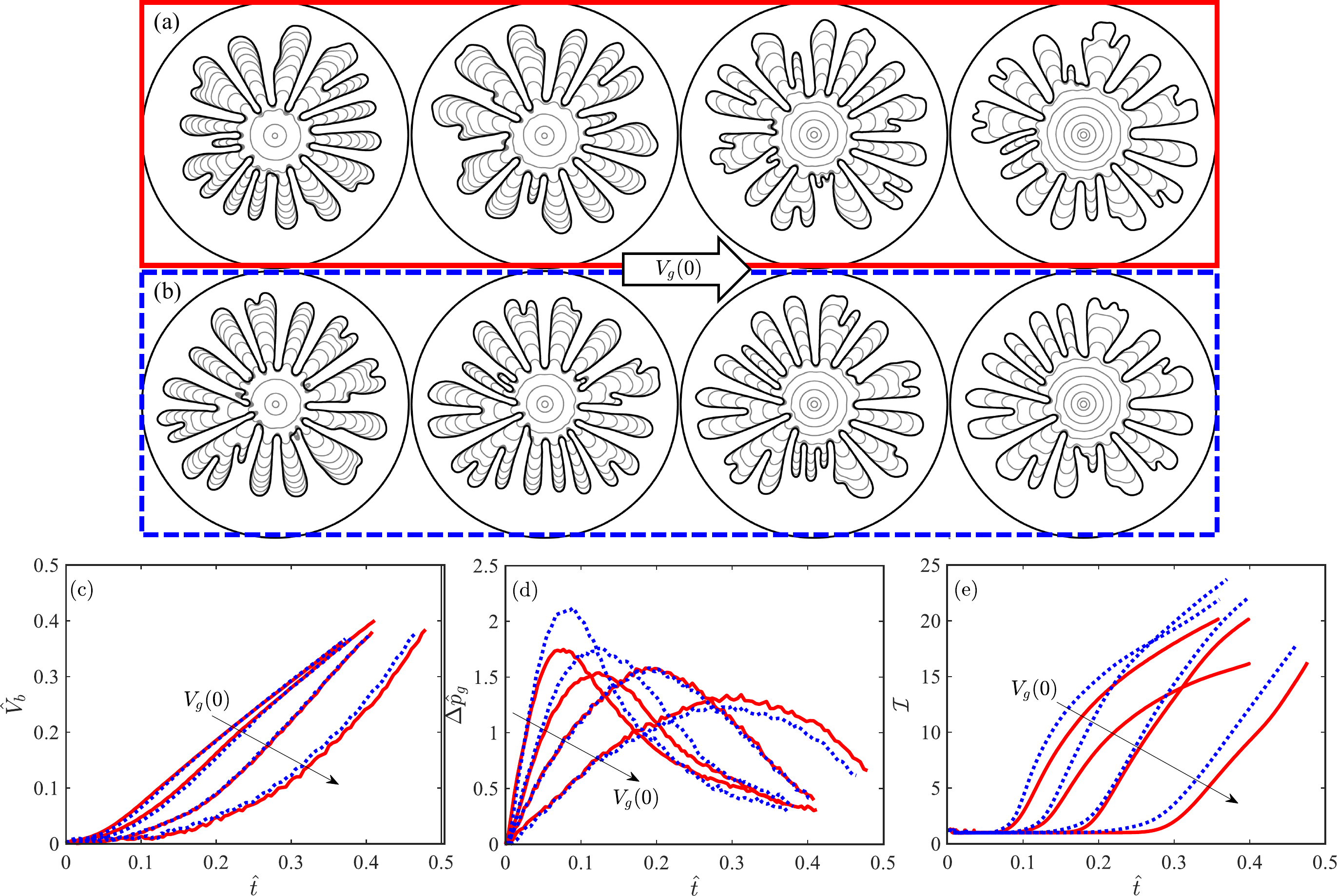}
	\caption{Experimental reproducibility. (a, b) Interface evolution for experimental repetitions at $Q=2.5$~mL/min and (left to right) $V_\res(0)\in\{25,50,100,200\}$~mL. The experiments in (a) correspond to the Main Set reported throughout the main document, while those in (b) are the Repeat Set. (c-e) Non-dimensional (c) volume, (d) gauge gas pressure $\Delta\hat{p}_g$, and (e) isoperimetric ratio as functions of non-dimensional time for the experiments shown in (a) and (b), plotted as solid red and dotted blue curves, respectively. Arrows indicate increasing $V_g(0)$.}
	\label{fig:reprod}
\end{figure}

Figure~\ref{fig:reprod} shows data from two sets of experiments performed at the same parameter values ($Q=2.5$~mL/min; $V_g(0)$=25, 50, 100, and 200~mL). The Main Set [Fig.~\ref{fig:reprod}(a)] corresponds to the set of data presented throughout the main text, i.e., in Figs.~\ref{fig:Experiments}, \ref{fig:VolPressureComp}(b, d), \ref{fig:metrics}(b). The Repeat Set [Fig.~\ref{fig:reprod}(b)] are used with the Main Set to calculate the mean values shown in Fig.~\ref{fig:metricsFin}, along with the experimental error bars.

Comparing the two sets, we see that they both exhibit the same qualitative response to increasing the initial gas volume $V_g(0)$ (arrows), which leads to the growth of fingers at larger radii. Quantitatively, the dynamical growth of the bubbles $\hat{V}_b(\hat{t})$ [Fig.~\ref{fig:reprod}(c)] is remarkably reproducible; data from the Main Set (solid red curves) and Repeat Set (dotted blue curves) largely overlap on the scale of the plot. This is despite significant differences in the evolution of both the gauge gas pressure $\Delta\hat{p}_g$ [Fig.~\ref{fig:reprod}(d)] and the viscous fingering pattern, again quantified by the isoperimetric ratio $\mathcal{I}(\hat{t})$ [Fig.~\ref{fig:reprod}(e)]. While the observable onset of fingering (when $\mathcal{I}>1.1$) is in reasonable agreement between the two sets, the Repeat Set has systematically greater $\mathcal{I}$ than the Main Set. The Hele-Shaw cell was deconstructed, cleaned and reassembled between the two sets, which will have introduced small variations in cell geometry between the sets and is likely the reason for the systematic offset in $\mathcal{I}(\hat{t})$. The greater variability in both $\Delta\hat{p}_g$ and $\mathcal{I}$ than in $\hat{V}_b$ may reflect the greater sensitivity of pressure to variations in fingering pattern (Fig.~\ref{fig:VolPressureCa}).

\end{document}